\documentclass[pra,twocolumn,showpacs,superscriptaddress,floatfix, reprint]{revtex4}

\usepackage[OT1]{fontenc}
\usepackage[usenames,dvipsnames]{color}
\usepackage[latin1]{inputenc}
\usepackage[english]{babel}

\usepackage{chngcntr}
\usepackage[toc,page]{appendix}

\usepackage{graphicx}
\usepackage{color}
\usepackage{mathtools}
\usepackage{amssymb,amsmath}
\usepackage[Gray,squaren]{SIunits}
\usepackage{xspace}
\usepackage{tabstackengine}
\stackMath
\usepackage{upgreek}
\usepackage{ulem}
\usepackage{epstopdf}
\usepackage{amssymb,amsmath,verbatim,ulem}
\usepackage{booktabs}

\newcommand{\ket}[1]{|{#1}\rangle}
\newcommand{\bra}[1]{\langle{#1}|}

\begin{document}

\title{Resolving closely spaced levels for Doppler mismatched double resonance}
\author{Elijah Ogaro Nyakang'o}
\email{eogaro@gmail.com}
\address{Department of Physics, Indian Institute of Technology Guwahati, Guwahati, Assam 781039, India}
\author{Kanhaiya Pandey}
\email{kanhaiyapandey@iitg.ac.in}
\address{Department of Physics, Indian Institute of Technology Guwahati, Guwahati, Assam 781039, India}

\date{\today{}}

\begin{abstract}
In this paper, we present experimental techniques to resolve the closely spaced hyperfine levels of a weak transition by eliminating the residual/partial two-photon Doppler broadening and cross-over resonances in a wavelength mismatched double resonance spectroscopy. The elimination of the partial Doppler broadening is based on velocity induced population oscillation (VIPO) and velocity selective saturation (VSS) effect followed by the subtraction of the broad background of the two-photon spectrum. Since the VIPO and VSS effect are the phenomena for near zero velocity group atoms, the subtraction gives rise to Doppler-free peaks and the closely spaced hyperfine levels of the $6\text{P}_{3/2}$ state in Rb are well resolved. The double resonance experiment is conducted on $5\text{S}_{1/2}\rightarrow5\text{P}_{3/2}$ strong transition (at 780~nm) and $5\text{S}_{1/2}\rightarrow6\text{P}_{3/2}$ weak transition (at 420~nm) at room temperature.   
\end{abstract}

%\pacs{42.50.Md, 42.25.Dd}

\maketitle

\section{Introduction}

Saturated absorption spectroscopy \cite{IJO01} is a commonly used technique in the field of laser spectroscopy to overcome the Doppler broadening effect by canceling it in the counter-propagation configuration of the probe and pump lasers. However, the drawback of this technique is the formation of spurious (or cross-over) resonance peaks within the spectrum peaks, which swamps the real resonance peaks if the levels are closely spaced within the Doppler profile.

Further, the cancellation of the Doppler effect for two-photon (or multi-photon) processes such as electromagnetically induced transparency (EIT) \cite{BIH91,LYX95} requires appropriate lasers propagation direction. However, this cancellation is only possible if the wavelength of the lasers is approximately the same \cite{DAN06,DPW06,LYX95,JYS95} otherwise suffers through partial Doppler broadening due to wavelengths mismatch of the transitions involved \cite{SFD96,BZF98,KPW05,MJA07}. Recently this mismatch has been recovered using velocity dependent light shift for detuned control laser and with an extra dressing laser \cite{FLM19, LFD19}.

In this work, we eliminate both of these problems, i.e. (i) the cross-over peaks formed within the spectrum peaks and (ii) wavelength mismatched partial Doppler broadening, for double resonance at 780~nm and 420~nm of a V-type system to resolve closely spaced hyperfine level in $^{85}$Rb. The blue transition ($5\text{S}_{1/2}\rightarrow6\text{P}_{3/2}$) at 420~nm is weak and the infrared (IR) transition ($5\text{S}_{1/2}\rightarrow5\text{P}_{3/2}$) at 780~nm is strong. The direct detection of absorption on the weak blue transition \cite{PBA15,GKK20} is a bit challenging and hence the double-resonance spectroscopy \cite{CAZ16,BZF98,PMH19,EDV20,ZLT14} is commonly used which again suffers through partial Doppler broadening. The previously double resonance spectroscopy at 420 nm and 780 nm in Rb was mainly done in $^{87}$Rb due to the limitation posed by the residual Doppler broadening effect \cite{CAZ16,BZF98,EDV20,ZLT14}. Resolving the hyperfine levels and stabilizing the blue laser at particular transition of Rb is very important for precision measurement \cite{NDH19,EDV20,GKK20} and laser cooling as the expected temperature is 5 times lower in the magneto-optical trap than the routinely used IR transition, similar to the case of K \cite{MJF11} and Li \cite{DHH11}. This transition is also useful for the coherent Rydberg excitation of Rb for quantum computation and information processing \cite{SAA17}.

The method used to overcome the above mentioned two problems are velocity induced population oscillation and velocity selective saturation (VSS) effects. In the atomic frame for the moving atoms, the two counter-propagating lasers with the same polarization and driving the same transition, will be beating due to opposite Doppler shift. The beating of the two lasers causes a temporal modulation of population difference between the levels driven by the lasers and the phenomenon is called population oscillation \cite{BBM05,RWB09,PRB07,ZKG06,BLB03,KSB18,MWR16}. Since the two beating fields have same frequency, the induced population oscillation is dependent on the velocity of the atom and hence the name velocity induced population oscillation (VIPO) \cite{OKP20}. The VIPO effect occurs only for a narrow range of beat frequencies (i.e. near zero velocity range) because of the inherent population inertia i.e. the slow response of electric dipoles to incident fields. The range of beat frequencies is determined by the inverse of population relaxation times of the upper levels \cite{HBK83,MWR16,OKP20}. Similarly, VSS effect is also for near zero velocity group atom and hence the effect of partial Doppler broadening and cross-over peaks is removed for multi-photon resonance.

This paper is organized in the following way. In section \ref{Setup}, we describe the relevant energy levels with the transitions of the various configurations and the experimental setup. In section \ref{Theory}, we describe the density matrix formalism for the various systems considered and the numerically simulated absorption profile of the probe. In section \ref{ExpRST}, we present the experimental results on resolving the closely spaced hyperfine levels of the $6\text{P}_{3/2}$ state in $^{85}$Rb and $^{87}$Rb. Finally in section \ref{Conc}, we give the conclusion on this work.

\section{Energy levels and Experimental set up}\label{Setup}

\begin{figure}
 \begin{center}    
 \includegraphics[width =1.0\linewidth]{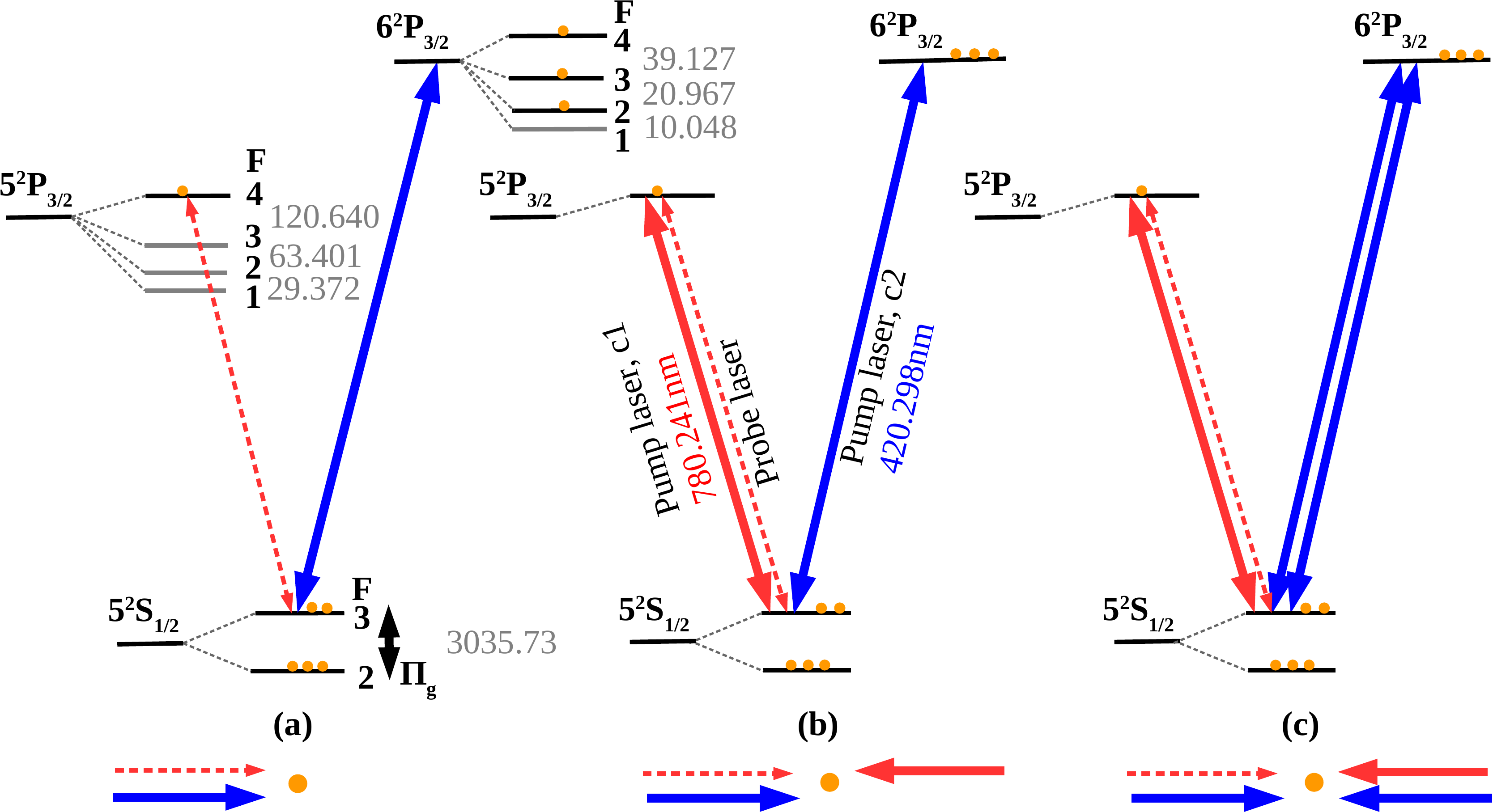}
     \caption{(Color online). Energy levels ($^{85}\text{Rb}$) with hyperfine splitting (in MHz) and the various transitions in different configurations for EIT. (a) V-type open system, (b) V-type open system with the VIPO effect at IR transition (c) V-type open system with the VIPO effect at IR transition and VSS effect at blue transition. $\Pi_{g}\approx 40~\text{kHz}$ \cite{LYX95} is the ground state mixing rate.}
    \label{Fig1}
   \end{center}
\end{figure}

\begin{figure}
  \begin{center}    
  \includegraphics[width =1.0\linewidth]{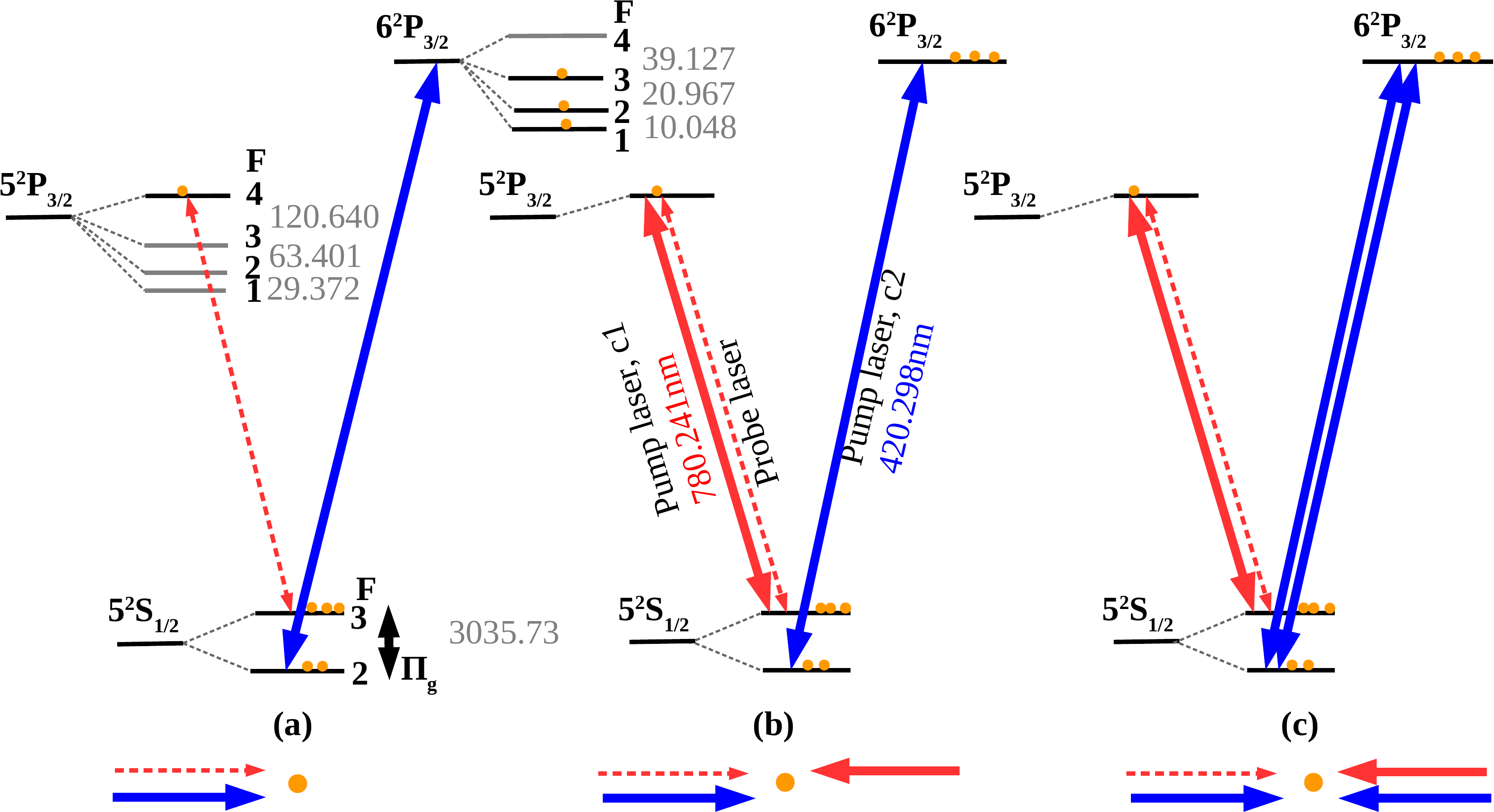}
     \caption{(Color online). Energy levels ($^{85}\text{Rb}$) with hyperfine splitting (in MHz) and the various transitions in different configurations for EA scheme. (a) Optical pumping system, (b) Optical pumping system with the VIPO effect at IR transition (c) Optical pumping system with the VIPO effect at IR transition and VSS effect at blue transition.}
     \label{Fig2}
   \end{center}
\end{figure}

The relevant energy levels and transitions is illustrated in Fig. \ref{Fig1} and \ref{Fig2} for the V-type system and optical pumping system respectively. The propagation direction of the probe and the pump (or control) lasers at 780~nm (IR) and 420~nm (blue) transitions in various configurations is shown below the energy level scheme.
%%%% Figure 1  %%%%%%%%%%%%%%%%%
%%%%%%%%%%%%%%%%%%%%%%%%%%%%%%%%%%%%%%%

%%%% Figure 2  %%%%%%%%%%%%%%%%%%%%%%%%
%%%%%%%%%%%%%%%%%%%%%%%%%%%%%%%%%%%%%%

The probe and the counter-propagating pump lasers at $780~\text{nm}$ are locked to resonance on $5\text{S}_{1/2}(\text{F}=3)\leftrightarrow5\text{P}_{3/2}(\text{F}=4)$ transition. The lifetime, $\uptau_{1}$ of the state, $5\text{P}_{3/2}(\text{F}=4)$ is $26.25~\text{ns}$ \cite{GAF02,SMS11,VUS96}. The absorption of the probe is monitored as the $420~\text{nm}$ pump laser scans across the $6\text{P}_{3/2}$ hyperfine levels on $5\text{S}_{1/2}(\text{F}=3)\leftrightarrow6\text{P}_{3/2}$ weak transition for a V-type system or on $5\text{S}_{1/2}(\text{F}=2)\leftrightarrow6\text{P}_{3/2}$ weak transition in  the case of optical pumping system. The  lifetime, $\uptau_{2}$ of $6\text{P}_{3/2}$ is $120.7~\text{ns}$ \cite{GAO04}.

The $780~\text{nm}$ laser beam is generated from the thorlab laser diode L785H1 which is a home-assembled extended cavity diode laser (ECDL) with typical linewidth of $500~\text{kHz}$. This laser is locked to resonance on $5\text{S}_{1/2}(\text{F}=3)\leftrightarrow5\text{P}_{3/2}(\text{F}=4)$ transition shown in Figs. \ref{Fig1} and \ref{Fig2} using saturated absorption spectroscopy (SAS) set-up. The error signal for locking the laser is generated by frequency modulation using the current of ECDL at $50~\text{kHz}$. The recorded experimental spectra is frequency scaled using the resolved peaks location of the green trace (in each of the configurations) for the hyperfine splitting values given in reference \cite{GKK20}.

The $420~\text{nm}$ laser beam is generated from a commercially available ECDL from TOPTICA of model no. DL PRO HP with a typical linewidth of $<200~\text{kHz}$ and output power of $70~\text{mW}$. A portion of the beam is fed to Fabry-Perot Interferometer for monitoring the single-mode operation of the blue laser. The beam diameter of the $780~\text{nm}$ probe and pump beams is $2\times 3~\text{mm}$ and that of $420~\text{nm}$ pump beams is $3\times4~\text{mm}$. The power of the probe beam used in the experiment is $42~\mu\text{W}$ (or peak intensity, $\text{I}=1.78~\text{mW}/\text{cm}^2$).

\begin{figure}
   \begin{center}    
    \includegraphics[width =1.0\linewidth]{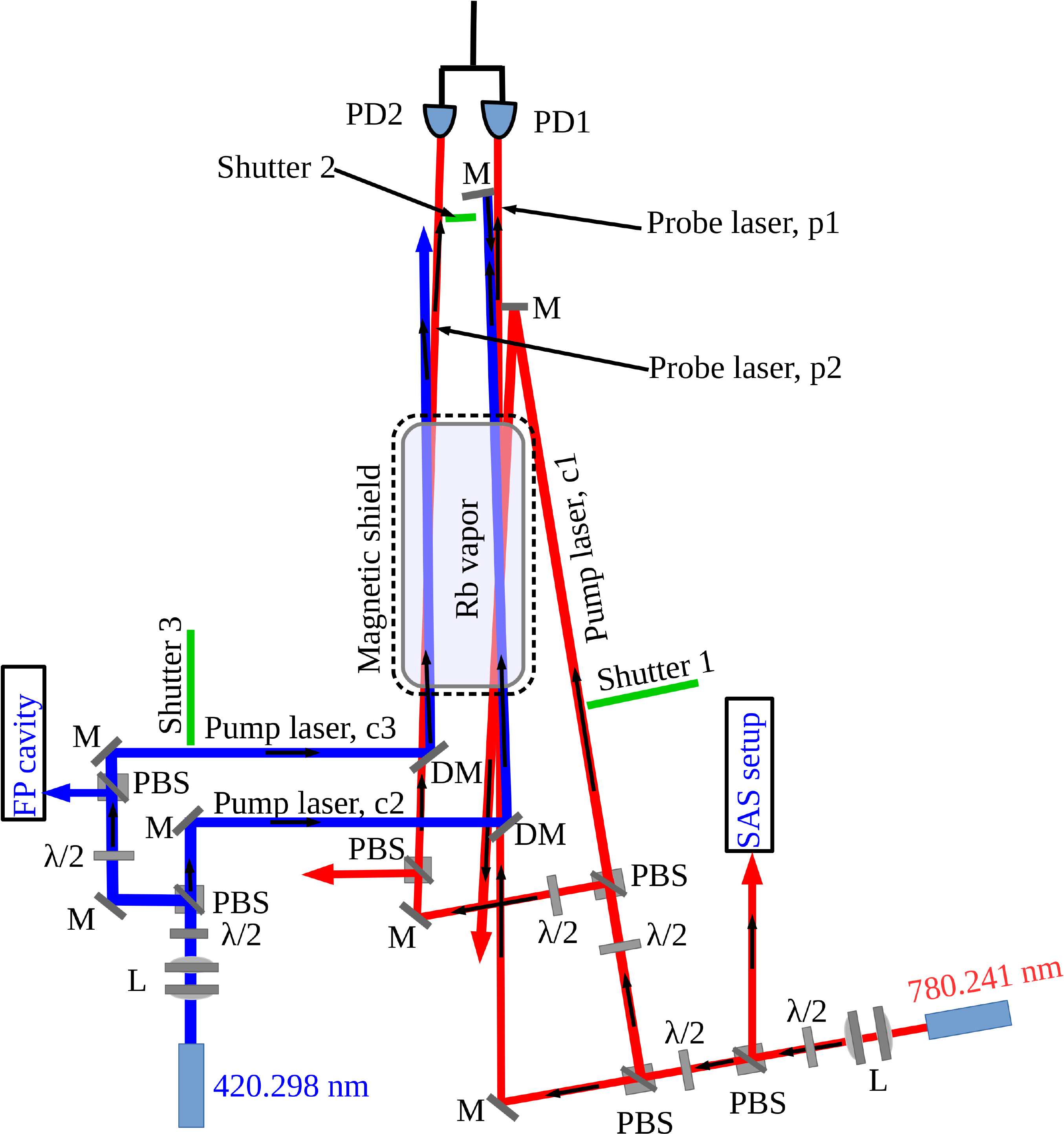}
      \caption{(Color online). The experimental setup for for resolving the closely spaced hyperfine levels of the $6\text{P}_{3/2}$ state in Rb atom using the VIPO and VSS effects.}
      \label{Fig3}
   \end{center}
\end{figure}

The detailed experimental set up is shown in Fig. \ref{Fig3}. In order to extract the narrow linewidth, the probe laser beam is divided into two beams with same polarization and power and propagated in the Rb cell with a spatial separation of about 1~cm. The blue beam is also divided into two beams with the same polarization as the IR beams and co-propagates with the two probes as shown in experimental set-up of Fig. \ref{Fig3}. The IR pump beam which counter-propagates with one of the probe beam, has the same polarization as the probe beam since having the same polarization is key for the interference/beating of the two fields. The interference/beating of the fields requires the polarization of the two fields to be identical and this aspect has been verified experimentally by rotating the polarization of one of the fields. When the polarization of the two fields are orthogonal, the VIPO dip disappear. There is a retro-mirror for reflecting the blue beam (which is overlapping with the IR pump beam) to generate counter-propagating blue beams in the cell when shutter 2 is open. It is very important to keep the angle between the beams as minimum as possible (i.e. near zero angle) and also use a magnetic shield to minimize broadening of the spectrum.

There are three shutters which are used to generate various conditions and configurations in the experiment. The configuration represented by  Fig. \ref{Fig1}a or \ref{Fig2}a is generated with all the shutters closed. The configuration represented by Fig. \ref{Fig1}b or \ref{Fig2}b is generated with shutter 1 open and shutter 2 closed. The configuration represented by Fig. \ref{Fig1}c or \ref{Fig2}c is generated with shutter 1 and shutter 2 open. Opening the shutter 3 removes the broad background of the transparency and EA peaks. The broad background is removed by the subtraction of the absorption/transparency spectra of the two probes using two identical IR photo-detectors (PD1 and PD2) in the differential transimpedance amplifier.
%%%% Figure 3  %%%%%%%%%%%%%%%%%%%% 
%%%%%%%%%%%%%%%%%%%%%%%%%%%%%%%%%%%%

\section{Theoretical model}\label{Theory}

We have conducted the experiments in the six configurations shown in Fig. \ref{Fig1} and \ref{Fig2}, three of them are for the V-type open system (Fig. \ref{Fig1}a, \ref{Fig1}b and \ref{Fig1}c) and the other three are for optical pumping system (Fig. \ref{Fig2}a, \ref{Fig2}b and \ref{Fig2}c). The V-type open system is further sub-categorized into: (i) V-type open system shown in \ref{Fig1}a, (ii) VIPO at IR transition for V-type open system  shown in Fig. \ref{Fig1}b and (iii) VIPO at IR and VSS at blue transition for V-type open system shown in Fig. \ref{Fig1}c. Similarly, the optical pumping system is sub-categorized into: (i) Optical pumping system shown in Fig. \ref{Fig2}a, (ii) VIPO at IR transition for optical pumping system shown in Fig. \ref{Fig2}b and (iii) VIPO at IR and VSS at blue transition for optical pumping system shown in Fig. \ref{Fig2}c. We discuss the theory for all these configurations one by one.   

\subsection{Transparency for V-type open system }

\subsubsection{V-type open system}
\label{V_system}

This corresponds to the energy level and the configuration shown in Fig. \ref{Fig1}a and is achieved by closing all the shutters of the experimental setup in Fig. \ref{Fig3}. This system is very well known and has been extensively studied \cite{BZF98,VBA07}. This V-type of system is open as the population from $6\text{P}_{3/2}$  decays to the other ground state hyperfine level, $5\text{S}_{1/2}(\text{F}=2)$ and can not be recycled. In the presence of the blue pump laser, c2, there is transparency of the IR probe laser due to two effects, one is coherence effect i.e. EIT in a V-type atomic system \cite{DAV05,MSA99} and the other is optical pumping effect \cite{FBK80,SMH04,HRN09}. The Hamiltonian of the system, the equations of motion and the analytical expression for the absorption of the probe, $\rho_{12}$, are given in Eq. \ref{eqA1}, \ref{eqA2} and \ref{eqA3} respectively.

The mixing rate, $\Pi_g$, for the hyperfine ground states (appearing in the equations of motion) is due to thermal collisions and the time of fight of atoms across the laser beam \cite{LYX95,SLR09}. The contribution due to time of flight is defined as $d/\tilde{v}$ where, $\tilde{v}$ is the thermal velocity of the atoms in the atomic medium and $d$ is the diameter of the laser beam. The numerically simulated absorption spectrum of the IR probe laser locked to resonance on $5\text{S}_{1/2}(\text{F}=3)\leftrightarrow5\text{P}_{3/2}(\text{F}=4)$ cycling transition vs detuning of the blue pump laser is plotted in Fig. \ref{Fig4} (see the blue trace). The Lorentzian fitting to this curve gives a linewidth of 16~MHz, while the linewidth is 11 MHz if the pump laser wavelength is taken to be 780~nm instead of 420~nm (see the table \ref{tab1}). This broadening by 1.5 times is due to residual or partial Doppler broadening caused by wavelength mismatch between the probe and the pump laser.

\subsubsection{VIPO at IR transition for V-type open system}
\label{V_system_SPO}

This configuration corresponds to the energy scheme given in Fig. \ref{Fig1}b and the experimental set-up when shutter 1 is open and shutter 2 is closed. This is theoretically modeled by considering the Hamiltonian $\text{H}$ under electric-dipole and rotating-wave approximation and in the interaction picture as follows,
\begin{align}
\label{eq1}
H=&\frac{\hbar}{2}\big\{(\Omega_\text{c1}+\Omega_\text{p}e^{i\delta_{1}{t}})\ket{1}\bra{2}+\Omega_\text{c2}\ket{1}\bra{3}\nonumber\\
&-\Delta_\text{c1}\ket{2}\bra{2}-\Delta_\text{c2}\ket{3}\bra{3}+h.c.\big\}
\end{align}
where, $5\text{S}_{1/2}(\text{F}=3)=\ket{1}$, $5\text{P}_{3/2}(\text{F}=4)=\ket{2}$, $6\text{P}_{3/2}(\text{F}=2)=\ket{3}$, $\delta_{1}=(\omega_\text{p}-k_{1}v)-(\omega_\text{c1}+k_{1}v)=-2k_{1}v$ is the frequency difference between IR probe and pump beams in the atomic frame (since $\omega_\text{p}=\omega_\text{c1}$), $k_{1}=2\pi/\lambda_{1}$ is the wave-vector of the IR laser and $\lambda_{1}$ is the wavelength, $v$ is the velocity of the atom in the direction of the probe, $\Delta_\text{c1}=\omega_\text{c1}-(\omega_{2}-\omega_{1})+k_{1}v$ is the detuning of the IR control laser, $\Delta_\text{c2}=\omega_\text{c2}-(\omega_{3}-\omega_{1})-k_{2}v$ is the detuning of the blue laser, $k_{2}=2\pi/\lambda_{2}$ is the wave-vector of the blue laser and $\lambda_{2}$ is the wavelength. The Rabi frequency for the fields is $\Omega_{\text{L}}=-\text{d}_{ij}\text{E}_{\text{L}}/\hbar$ where, $\text{d}_{ij}=\bra{i}\hat{\text{d}}\ket{j}$ is the dipole matrix element, $\hat{\text{d}}$ is the atomic dipole operator and subscript $\text{L}=\text{p},\text{c1},\text{c2}$ represent the fields (i.e. p is the probe and c1 is the pump of the $780~\text{nm}$ laser and c2 is the pump of the $420~\text{nm}$ laser).

The atom-field interaction is described by writing the Liouville-von Neumann equation for the density matrix,
\begin{align}
\label{eq2}
&\dot{\rho}=-\frac{i}{\hbar}[\text{H},\rho]-\frac{1}{2}\{\Gamma,\rho\}
\end{align}
where, $\rho$ is the atomic density operator, $\Gamma$ is the relaxation operator defined as $\bra{i}\Gamma\ket{j}=\gamma_{i}\delta_{ij} ~(\delta_{ij}=1$ if $i=j$ and 0 if $i\neq{j})$ and $\gamma_{i}$ is the decay rate of state $\ket{i}$. The temporal behavior of the element of density matrix governed by Eq. \ref{eq2} is velocity dependent due to the Doppler effect and oscillates at the harmonics of the beat frequency $\delta_{1}=-2k_{1}v$. The oscillation is caused by the beating of the two fields addressing the same transition $5\text{S}_{1/2}(\text{F}=3)\leftrightarrow5\text{P}_{3/2}(\text{F}=4)$ in Fig. \ref{Fig1}a. The equations of motion of the density matrix elements is given in Eq. \ref{eqA4} and is obtained using Eq. \ref{eq1} and \ref{eq2}. The harmonically oscillating density matrix elements at beat frequency can be written in the Floquet expansion \cite{SJH65,FZS05,UHH19} in the following form
\begin{align}
\label{eq3}
\rho_{ij}(t)=&\sum_{n=-\infty}^{\infty}\rho^{(n)}_{ij}(t)e^{in\delta_{1}{t}}
\end{align}
where, $\rho^{(n)}_{ij}(t)$ are $n^{\textrm{th}}$ harmonic amplitudes of the density matrix elements. The imaginary part of the zeroth harmonic, $\rho^{(0)}_{12}$ corresponds to the IR pump absorption, while the imaginary part of the first harmonic, $\rho^{(+1)}_{12}$ is for IR probe absorption in first order and all the others are for wave-mixing \cite{BRN81}. In the steady state condition ($\dot{\rho}^{(n)}_{ij}=0$ for all $n, i$ and $j$), the absorption of the probe laser ($\rho^{(+1)}_{12}$) is obtained by substituting the truncated series of the Floquet expansion given in Eq. \ref{eq3} up to first-order into Eq. \ref{eqA4}. The coefficients of the same power in $n\delta_{1}$ are then compared which yields a set of steady state  equations of motion in the Floquet expansion. The $\rho^{(+1)}_{12}$ element of the density matrix is expressed as follows, 
\begin{align}
\label{eq4}
\rho^{(+1)}_{12}=&\overbrace{\frac{i\Omega_{p}}{2(\gamma_{12}+i\delta_{1})}(\rho^{(0)}_{11}-\rho^{(0)}_{22})}^{\text{I}}+\nonumber\\
&\underbrace{\frac{i\Omega_{c1}}{2(\gamma_{12}+i\delta_{1})}(\rho^{(+1)}_{11}-\rho^{(+1)}_{22})}_\text{II}-\underbrace{\frac{i\Omega_{c2}}{2(\gamma_{12}+i\delta_{1})}\rho^{(+1)}_{32}}_\text{III}\nonumber\\
\end{align}
where, $\gamma_{12}=i\Delta_{p}+\gamma^{dec}_{12}$, $\Delta_{p}=\Delta_{c1}=0$, $\gamma^{dec}_{ij}={\frac{1}{2}}(\Gamma_{i}+\Gamma_{j})$ and $\Gamma_{i}$ is the decay rate of the $i^{\text{th}}$ level. The quantity $(\rho^{(0)}_{11}-\rho^{(0)}_{22})$ in term I is the population inversion created by the pump lasers at IR and blue transition. The quantity $(\rho^{(+1)}_{11}-\rho^{(+1)}_{22})$ in term II is the population oscillation difference and its contribution is significant for the velocity group atoms in the range of $|v|\sim\Gamma_{2}/k_1$ and forms a dip inside the transparency window. The density matrix element $\rho^{(+1)}_{32}$ in term III is the coherence oscillation which further modifies the lineshape of the dip inside the transparency window. The role of individual terms for the probe absorption is shown in Fig. \ref{FigA2}.

\begin{figure}
   \begin{center}   
   \includegraphics[width =1.0\linewidth]{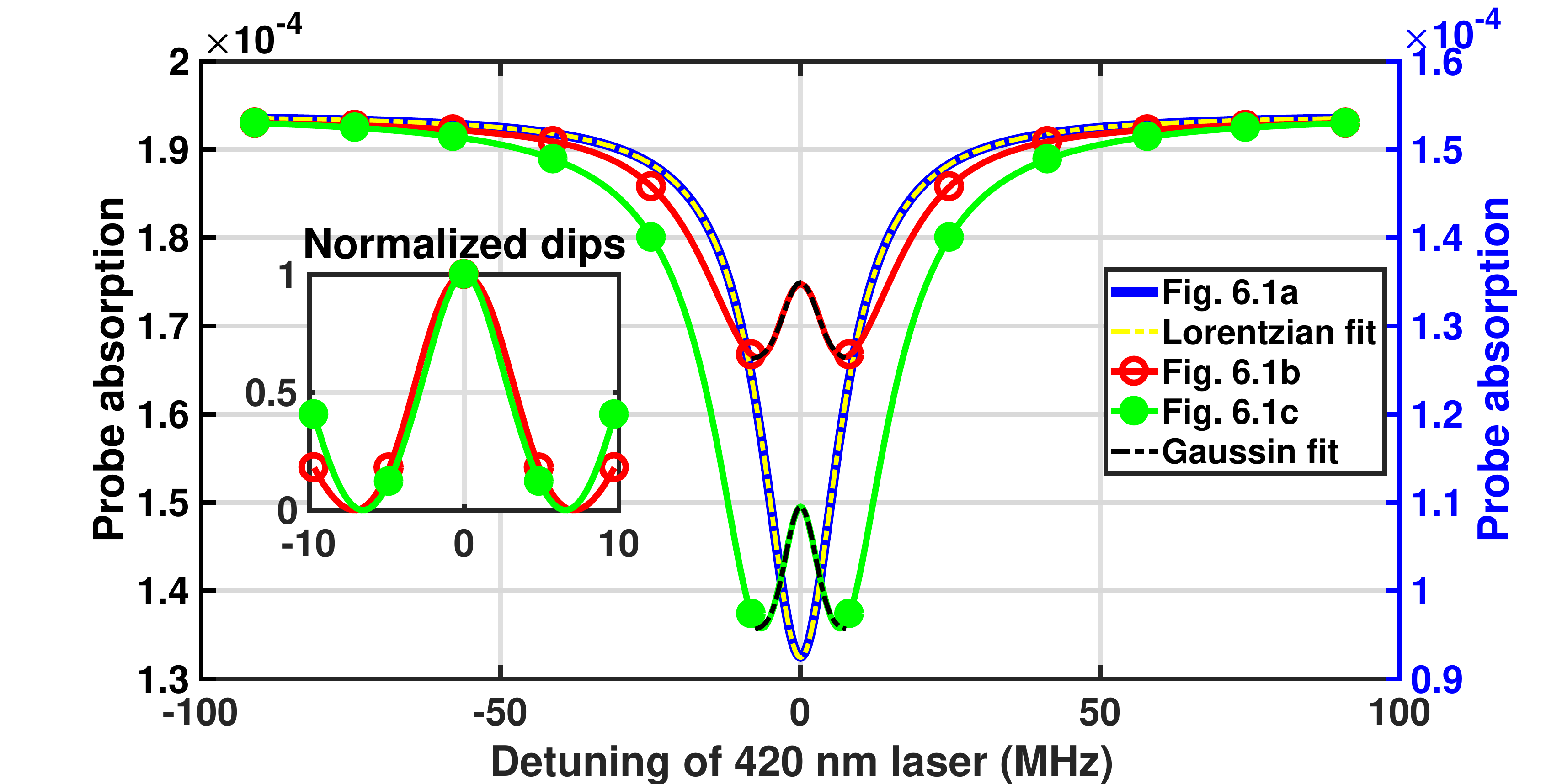}
      \caption{(Color online). Numerically calculated thermal averaged probe absorption vs detuning of 420~nm pump laser for V-type open system with various configurations shown in Fig. \ref{Fig1} (where $\Omega_{p}=\sqrt{0.0005}\Gamma_2$). The blue curve corresponds to Fig. \ref{Fig1}a with $\Omega_{c2}=\sqrt{1.5}\Gamma_3$. The red curve marked by circles corresponds to Fig. \ref{Fig1}b with $\Omega_{c1}=\Gamma_2$ and $\Omega_{c2}=\sqrt{1.5}\Gamma_3$. The green curve marked with dots corresponds to Fig. \ref{Fig1}c with $\Omega_{c1}=\sqrt{0.5}\Gamma_2$, $\Omega_{c2}=\sqrt{1.5}\Gamma_3$, $\Gamma_{2}=2\pi\times6.065~\text{MHz}$ and $\Gamma_{3}=2\pi\times1.32~\text{MHz}$. The vertical axis of the blue trace is on the left and the red trace marked by circles and the green trace marked with dots are on the right.}
      \label{Fig4}
   \end{center}
\end{figure}

The absorption of the probe laser is obtained by thermal averaging of Eq. \ref{eq4} at room temperature as follows, $\frac{1}{\sqrt{2\pi}\tilde{v}}\int \rho^{(+1)}_{12} e^{-({\frac{v}{2\tilde{v}}})^2}\mathrm{d}v$, where, $\tilde{v}=\sqrt{k_{B}T/{m}}$, $m$ (= 85 a.m.u) is the atomic mass and $T$ (= 300~K) is the temperature. The lineshape of the probe absorption after thermal averaging is shown in Fig. \ref{Fig4} (see the red trace marked by circles). The linewidth of the dip inside the transparency window is around 7~MHz which is less than the linewidth for a V system if the pump laser had wavelength at 780~nm instead of 420~nm. The linewidth of the dip is determined by fitting with a Gaussian line-profile (which fits better than a Lorentzian line-profile). The FWHM of a Gaussian fit (i.e. A$e^{-(x-x_{0})^2/(2\sigma^2)}$), is $2\sqrt{2\ln{2}}\sigma$ where A, $x_{0}$ and $\sigma$ are the fitting parameters and x is the frequency detuning. 
%%%% Figure 4  %%%%%%%%%%%%%%%%%%%%%%%
%%%%%%%%%%%%%%%%%%%%%%%%%%%%%%%%%%%%%%%
%%%%% Table 1  %%%%%%%%%%%%%%%%%%%%%%%%
\begin{table}[ht]
\begin{center}
\caption{Comparison of numerically calculated linewidth for various configuration}
\label{tab1}
\begin{tabular}{ |p{7cm}|p{1.43cm}|  }
 \hline
 System and configuration & Linewidth (MHz)  \\
 \hline
Configuration as shown in Fig. \ref{Fig1}a  & 16\\ 
Configuration as shown in Fig. \ref{Fig1}a but considering pump wavelength 780~nm instead of 420~nm & 11 \\
 \hline
Configuration as shown in Fig. \ref{Fig1}b &   7$\pm1$ \\ 
Configuration as shown in Fig. \ref{Fig1}c &   6$\pm1$ \\
 \hline\hline 
Configuration as shown in Fig. \ref{Fig2}a&  17 \\  
Configuration as shown in Fig. \ref{Fig2}a but considering pump wavelength 780~nm instead of 420~nm &  12$\pm1$ \\ 
 \hline
 Configuration as shown in Fig. \ref{Fig2}b  &   9$\pm1$ \\ 
 Configuration as shown in Fig. \ref{Fig2}c&   6$\pm1$ \\
\hline
\end{tabular}
\end{center}
\end{table}
%%%%%%%%%%%%%%%%%%%%%%%%%%%%%%%%%%%%%%%%%%%

\subsubsection{VIPO at IR and VSS at blue transition for V-type open system}
\label{V_system_DPO}

The energy scheme for this configuration is given in Fig. \ref{Fig1}c where the probe and IR pump are similarly locked to resonance on $5\text{S}_{1/2}(\text{F}=3)\leftrightarrow5\text{P}_{3/2}(\text{F}=4)$ cycling transition. The blue pump scans across the hyperfine levels of the $6\text{P}_{3/2}$ at the weak transition $5\text{S}_{1/2}(\text{F}=3)\leftrightarrow6\text{P}_{3/2}$ and is retro-reflected by mirror M to generate the two counter-propagating beams inside the Rb vapor cell. The VIPO on $5\text{S}_{1/2}(\text{F}=3)\leftrightarrow5\text{P}_{3/2}(\text{F}=4)$ transition will induce a dip on the transparency peak as previously explained in the Sec. \ref{V_system_SPO}. This dip is further enhanced by the VSS effect of the two counter-propagating blue pump laser beams.

The VSS effect can be understood in the following simple way. We consider population dynamics between the two states, $\ket{1}$ ($5\text{S}_{1/2}$, F=2)  and $\ket{3}$ ($6\text{P}_{3/2}$) due to two counter-propagating blue pump laser beams only in the absence of the IR laser. For simplicity, consider three velocity group of atoms, $+v$, $0$ and $-v$. For detuned case of the blue pump laser, ($\Delta_{c2}$) both the non-zero velocity group of atom $\pm v=\Delta_{c2}/k_2$ will be in resonance with either of the two counter-propagating blue pump laser and hence the number of atoms in the excited state will be twice. For zero detuning case, the near-zero ($<\Gamma_3/k_2$) velocity group of atom will be in resonance with both the blue pump laser and hence intensity seen by this group of atoms will be twice. However, the excited state population will be less than twice due to saturation effect, thus inducing a dip on the absorption spectrum of the probe beam with the scan of the blue pump laser.  The linewidth of this dip is in the range of $\Gamma_3$. This qualitative picture is also presented in \cite{SHY95}. Mathematically, the population transfer due to blue pump lasers will be given by the following equation \cite{OKP20}. 
\begin{align}
\label{eq5}
\rho_{33}&=\frac{1}{2}\Bigg\{\frac{\text{S}_{\Delta_{c2}-k_2v}+\text{S}_{\Delta_{c2}+k_2v}}{1+\text{S}_{\Delta_{c2}-k_2v}+\text{S}_{\Delta_{C2}+k_2v}}\Bigg\}
\end{align}
\normalsize{}with,\small{}
\begin{align}
\text{S}_{\Delta_{c2}+k_2v}&=\frac{\text{S}_0}{1+\frac{4(\Delta_{c2}+k_2v)^2}{\Gamma_3^2}},\enspace
\text{S}_{\Delta_{c2}-k_2v}=\frac{\text{S}_0}{1+\frac{4(\Delta_{c2}-k_2v)^2}{\Gamma_3^2}}\nonumber
\end{align}
\normalsize{}where $\text{S}_0(=2\Omega_{c2}^2/\Gamma_3)$ is the saturation intensity of the blue transition for the stationary atoms. In the presence of the IR pump laser i.e. when shutter 1 and 2 are open, the VSS effect will induce a dip on both the transparency spectra of both the IR pump and the probe.

The detailed formalism for the VIPO at IR and VSS at blue transitions is as follows. For the given velocity $v$ there is a beating for the two-counter-propagating blue pump lasers in the atomic frame with a beat frequency ($\delta_{2}=-2k_{2}v$). The Hamiltonian $\text{H}$ of a V-type system shown in Fig. \ref{Fig1}c under electric-dipole and rotating-wave approximation and in the interaction picture is as follows,
\begin{align}
\label{eq6}
\text{H}=&\frac{\hbar}{2}\big\{(\Omega_\text{c1}+\Omega_\text{p}e^{i\delta_{1}{t}})\ket{1}\bra{2}+(\Omega_\text{c2}+\Omega_\text{c2}e^{i\delta_{2}{t}})\ket{1}\bra{3}\nonumber\\
&-\Delta_\text{c1}\ket{2}\bra{2}-\Delta_\text{c2}\ket{3}\bra{3}+\text{h.c.}\big\}
\end{align}
The equations of motion of the density matrix elements is given in Eq. \ref{eqA5} which is obtained using Eq. \ref{eq2} and \ref{eq6}. The coefficients of the harmonically oscillating density matrix elements have two different time dependencies, which is also the case for the Hamiltonian in Eq. \ref{eq6}. The Floquet expansion for the density matrix elements in such a case is modified and written as follows,  
\begin{align}
\label{eq7}
\rho_{ij}(t)=&\sum_{m=-\infty}^{\infty}\Big(\sum_{n=-\infty}^{\infty}\rho^{(n,m)}_{ij}(t)e^{i(n\delta_{1}+m\delta_{2}){t}}\Big)
\end{align}
where, $n$ is the $n^{\text{th}}$ harmonic component due the beating of the IR laser beams and $m$ is the $m^{\text{th}}$ harmonic component due the beating of the blue pump laser beams. The imaginary part of $\rho^{(0,0)}_{12}$ corresponds to the IR pump absorption, while the imaginary part of $\rho^{(+1,0)}_{12}$ is for IR probe absorption and all the others are for wave-mixing. In the steady state condition (i.e. $\dot{\rho}^{(n,m)}_{ij}=0$ for all $n$, $m$, $i$ and $j$), $\rho^{(+1,0)}_{12}$ is obtained by substituting the truncated series of the Floquet expansion given in Eq. \ref{eq7} up to first-order into Eq. \ref{eqA5}. The coefficients of the same power in $(n\delta_{1},m\delta_{2})$ are similarly compared and yields a set of steady state equations of motion in the Floquet expansion. The $\rho^{(+1,0)}_{12}$ element of the density matrix is expressed as follows, 
\begin{align}
\label{eq8}
&\rho^{(+1,0)}_{12}={\Big\{\overbrace{\frac{i\Omega_{p}(\rho^{(0,0)}_{11}-\rho^{(0,0)}_{22})}{2(\gamma_{12}+i\delta_{1})}}^\text{I}}+\nonumber\\
&\underbrace{\frac{i\Omega_{c1}(\rho^{(+1,0)}_{11}-\rho^{(+1,0)}_{22})}{2(\gamma_{12}+i\delta_{1})}}_\text{II}-\underbrace{\frac{i\Omega_{c2}(\rho^{(+1,0)}_{32}+\rho^{(+1,-1)}_{32})}{2(\gamma_{12}+i\delta_{1})}}_{\text{III}}\Big\} \nonumber\\ 
\end{align}

In Eq. \ref{eq8}, the quantity $(\rho^{(0,0)}_{11}-\rho^{(0,0)}_{22})$ in term I is the population inversion induced by the IR pump and the saturation of the counter-propagating blue pumps and $(\rho^{(+1,0)}_{11}-\rho^{(+1,0)}_{22})$ in term II is the population oscillation induced by the beating of the IR probe and pump laser beams and the saturation of the counter-propagating blue pumps. The density matrix elements, $\rho^{(+1,0)}_{32}$  and $\rho^{(+1,-1)}_{32}$ in  term III, are the oscillating coherence terms due to the beating of the fields on IR and blue transitions. The thermal averaged probe absorption, $\frac{1}{\sqrt{2\pi}\tilde{v}}\int \rho^{(+1,0)}_{12} e^{-({\frac{v}{2\tilde{v}}})^2}\mathrm{d}v$ is calculated numerically and is plotted in Fig. \ref{Fig4} (see the green trace marked with dots). The linewidth of the induced dip is around 6~MHz.

\subsection{Enhanced absorption for optical pumping system}

\subsubsection{Optical pumping system}
\label{Opt_Pmp_EA}

This system corresponds to the energy level and the configuration shown in Fig. \ref{Fig2}a and is achieved when all the shutters in the experimental setup of Fig. \ref{Fig3} are closed. Again, the probe laser is locked to resonance on $5\text{S}_{1/2}(\text{F}=3)\leftrightarrow5\text{P}_{3/2}(\text{F}=4)$ transition. The absorption of the probe is monitored as the co-propagating blue pump laser scans across the $6\text{P}_{3/2}$ hyperfine levels on $5\text{S}_{1/2}(\text{F}=2)\rightarrow6\text{P}_{3/2}$ transition instead of $5\text{S}_{1/2}(\text{F}=3)\rightarrow6\text{P}_{3/2}$ transition. The absorption of the probe is increased by optical pumping of population to the upper ground hyperfine level $5\text{S}_{1/2}(\text{F}=3)$ \cite{FBK80,SMH04,HRN09} via $5\text{S}_{1/2}(\text{F}=2)\rightarrow6\text{P}_{3/2}(\text{F}=1,2,3)$ excitation and various decay channels (i.e. direct, $6\text{P}_{3/2}(\text{F}=2,3)\rightarrow5\text{S}_{1/2}(\text{F}=3)$ and indirect decay channels \cite{RMS12} such as $6\text{P}_{3/2}(\text{F}=1)\rightarrow6\text{S}_{1/2}\rightarrow5\text{P}_{3/2}\rightarrow5\text{S}_{1/2}(\text{F}=3)$). Therefore, optical pumping \cite{FSM95,TSR10} gives rise to enhanced absorption (EA) Doppler-free peaks of the $6\text{P}_{3/2}$ hyperfine levels. The numerically simulated absorption spectrum considering only one hyperfine level is plotted in Fig. \ref{Fig5} (see the blue trace). Note that the Hamiltonian of the system, equations of motion and the analytical expression for the absorption of the probe, $\rho_{12}$, are given in Eq. \ref{eqB1}, \ref{eqB2} and \ref{eqB3} respectively. The Lorentzian fitting to this curve gives a linewidth of 17~MHz, while it is 11 MHz if we consider the pump laser wavelength to be 780~nm instead of 420~nm (see the table \ref{tab1}). This broadening by 1.5 times is again due to residual or partial Doppler broadening caused by wavelength mismatch between the probe and the pump laser.

\subsubsection{VIPO at IR transition for optical pumping system}

This corresponds to the energy level and the configuration shown in Fig. \ref{Fig2}b and is achieved when shutter 2 is closed in the experimental setup of Fig. \ref{Fig3}. This is theoretically modeled by considering the Hamiltonian $\text{H}$ of the optical pumping system shown in Fig. \ref{Fig2}b under electric-dipole and rotating-wave approximation and in the interaction picture as follows,
\begin{align}
\label{eq9}
\text{H}=&\frac{\hbar}{2}\big\{(\Omega_\text{c1}+\Omega_\text{p}e^{i\delta_{1}{t}})\ket{1}\bra{2}+\Omega_\text{c2}\ket{4}\bra{3}\nonumber\\
&-\Delta_\text{c1}\ket{2}\bra{2}-\Delta_\text{c2}\ket{3}\bra{3}+h.c.\big\}
\end{align}
where, $5\text{S}_{1/2}(\text{F}=3)=\ket{1}$, $5\text{P}_{3/2}(\text{F}=4)=\ket{2}$, $6\text{P}_{3/2}(\text{F}=1)=\ket{3}$ and $5\text{S}_{1/2}(\text{F}=2)=\ket{4}$. The Hamiltonian is time dependent and the equations of motion of the density matrix elements is given in Eq. \ref{eqB4}. The equations of motion are solved in steady state after the Floquet expansion given in Eq. \ref{eq3} and the imaginary part of the density matrix element $\rho^{(+1)}_{12}$ gives the absorption of the probe as follows,
\begin{align}
\label{eq10}
&\rho^{(+1)}_{12}= \nonumber\\
&{\overbrace{\frac{i\Omega_{p}}{2(\gamma_{12}+i\delta_{1})}(\rho^{(0)}_{11}-\rho^{(0)}_{22})}^\text{I}+}\underbrace{\frac{i\Omega_{c1}}{2(\gamma_{12}+i\delta_{1})}(\rho^{(+1)}_{11}-\rho^{(+1)}_{22})}_\text{II}
\end{align}

The Eq. \ref{eq10} is similar to the Eq. \ref{eq4} except the coherence term. The first term, I in Eq. \ref{eq10} is due to population inversion created by the pump laser at IR and blue transition and gives only the EA line-shape. The second term is due to VIPO at IR transition and gives a dip inside the EA spectrum as shown Fig. \ref{Fig5} (see the red trace marked by circles). The linewidth of the dip is 9~MHz using Gaussian line profile fit. The contribution of each of the terms I and II is given in Fig. \ref{FigB2}.

\subsubsection{VIPO at IR and VSS at blue transition for optical pumping system}

The energy levels for this configuration is given in Fig. \ref{Fig2}c. The probe and the IR pump lasers are again locked to resonance on $5\text{S}_{1/2}(\text{F}=3)\leftrightarrow5\text{P}_{3/2}(\text{F}=4)$ cycling transition. The blue pump laser is scanning across the hyperfine levels of $6\text{P}_{3/2}$ at the weak transition, $5\text{S}_{1/2}(\text{F}=2)\leftrightarrow6\text{P}_{3/2}$ and is retro-reflected to generate the two counter-propagating beams inside the Rb vapor cell.

The Hamiltonian $\text{H}$ of the optical pumping system shown in Fig. \ref{Fig2}c under electric-dipole and rotating-wave approximation and in the interaction picture is given as follows,
\begin{align}
\label{eq11}
\text{H}=&\frac{\hbar}{2}\big\{(\Omega_\text{c1}+\Omega_\text{p}e^{i\delta_{1}{t}})\ket{1}\bra{2}+(\Omega_\text{c2}+\Omega_\text{c2}e^{i\delta_{2}{t}})\ket{4}\bra{3}\nonumber\\
&-\Delta_\text{c1}\ket{2}\bra{2}-\Delta_\text{c2}\ket{3}\bra{3}+h.c.\big\}
\end{align}
The probe absorption is similarly obtained in the steady state condition using the equations of motion given in Eq. \ref{eqB5} and the Floquet expansion given in Eq. \ref{eq7}. The imaginary part of the density matrix element $\rho^{(+1,0)}_{12}$ in the Floquet expansion gives the probe absorption and is expressed as follows,
\begin{align}
\label{eq12}
\rho^{(+1,0)}_{12}=&{\overbrace{\frac{i\Omega_{p}(\rho^{(0,0)}_{11}-\rho^{(0,0)}_{22})}{2(\gamma_{12}+i\delta_{1})}}^\text{I}}+\underbrace{\frac{i\Omega_{c1}(\rho^{(+1,0)}_{11}-\rho^{(+1,0)}_{22})}{2(\gamma_{12}+i\delta_{1})}}_\text{II}
\end{align}

\begin{figure}
   \begin{center}    
    \includegraphics[width =1.0\linewidth]{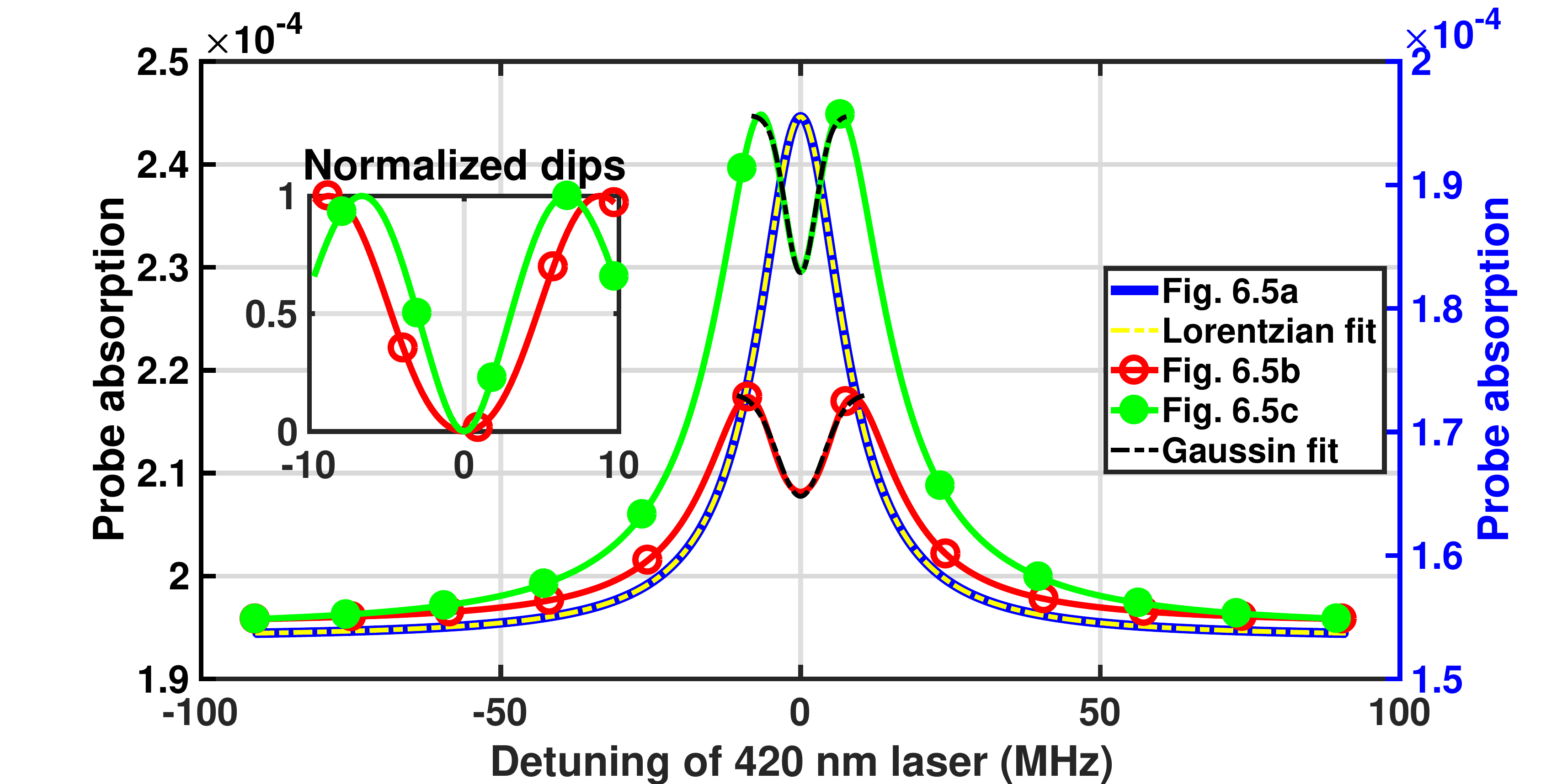}
      \caption{(Color online). Numerically calculated thermal averaged probe absorption vs detuning of 420~nm pump laser for optical system with various configurations shown in Fig. \ref{Fig2} (where $\Omega_{p}=\sqrt{0.0005}\Gamma_2$). The blue curve corresponds to Fig. \ref{Fig2}a with $\Omega_{c2}=\sqrt{1.5}\Gamma_3$. The red curve marked by circles corresponds to Fig. \ref{Fig2}b with $\Omega_{c1}=\Gamma_2$ and $\Omega_{c2}=\sqrt{1.5}\Gamma_3$. The green curve marked with dots corresponds to Fig. \ref{Fig2}c with $\Omega_{c1}=\sqrt{0.5}\Gamma_2$, $\Omega_{c2}=\sqrt{1.5}\Gamma_3$, $\Gamma_{2}=2\pi\times6.065~\text{MHz}$ and $\Gamma_{3}=2\pi\times1.32~\text{MHz}$. The vertical axis of the blue trace is on left and the red trace marked by circles and the green trace marked with dots are on the right.}
      \label{Fig5}
   \end{center}
\end{figure}

In Eq. \ref{eq12}, the quantity $(\rho^{(0,0)}_{11}-\rho^{(0,0)}_{22})$ in term I is the population inversion induced by the $780~\text{nm}$ and $420~\text{nm}$ pump lasers. The quantity $(\rho^{(+1,0)}_{11}-\rho^{(+1,0)}_{22})$ in term II is the population oscillation induced by the beating of the $780~\text{nm}$ laser beams and saturation effect induced by the counter-propagating $420~\text{nm}$ pump beams. The thermal averaged absorption in this configuration is shown in Fig. \ref{Fig5} (see the green trace marked with dots). The linewidth of the induced dip on the EA peak is about 6 MHz.
%%%%  Figure 5  %%%%%%%%%%%%%%%%%%%%%%%
%%%%%%%%%%%%%%%%%%%%%%%%%%%%%%%%%%%%%

\section{Experimental results}\label{ExpRST}
\subsection{Resolving $6\text{P}_{3/2}$ hyperfine levels in $^{85}\text{Rb}$}

\subsubsection{V-type open system}

\begin{figure}
  \begin{center}    
  \includegraphics[width =1.0\linewidth]{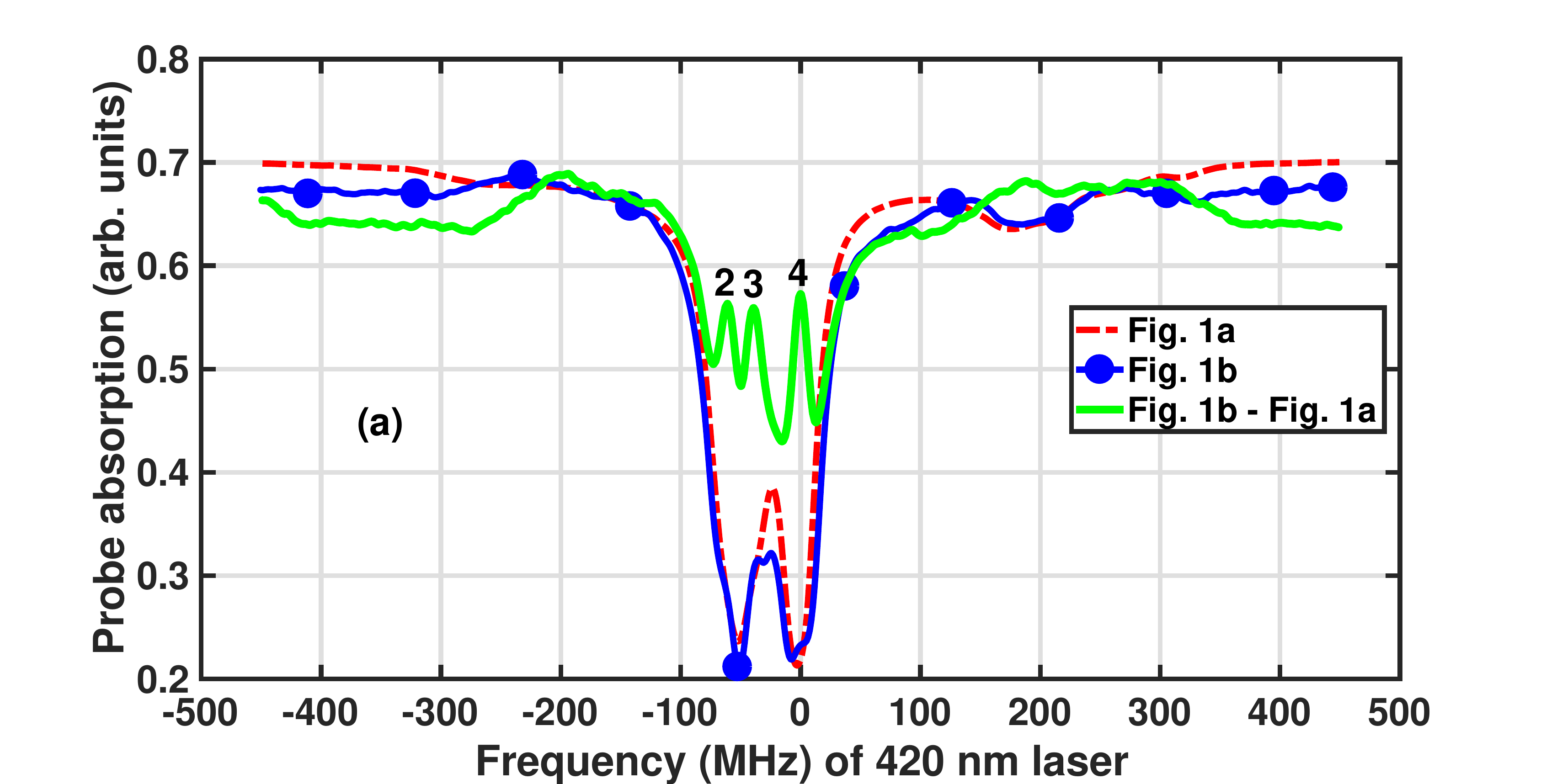}   
  \includegraphics[width =1.0\linewidth]{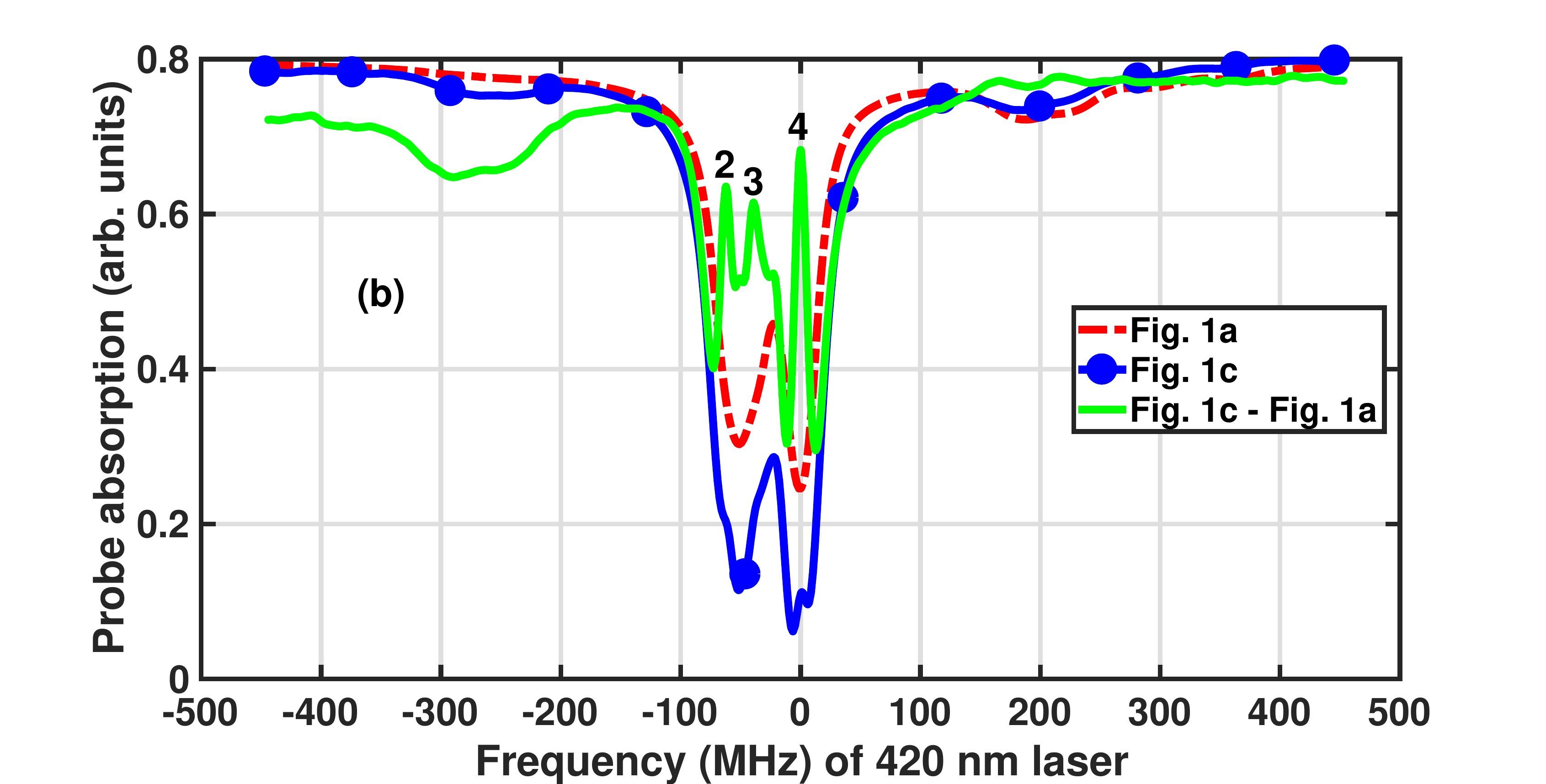}
    \caption{(Color online). The transparency spectrum of the $6\text{P}_{3/2}$ hyperfine levels in $^{85}\text{Rb}$ under various configurations shown in Fig. \ref{Fig1}. The red dashed trace is for the V-type open system (Fig. \ref{Fig1}a), the blue trace marked with dots in Fig. \ref{Fig6}a is for the V-type open system with VIPO effect at IR transition (Fig. \ref{Fig1}b) while the blue trace marked with dots in Fig. \ref{Fig6}b is for V-type open system with VIPO effect at IR and VSS effect at blue transition (Fig. \ref{Fig1}c). The green trace is the final result after removing the broad transparency background and it is magnified 3 times for visibility purpose.}
      \label{Fig6}
   \end{center}
\end{figure}

The transparency spectrum of the energy scheme in Fig. \ref{Fig1}a is shown by the red dashed trace of Fig. \ref{Fig6}. This spectrum is obtained when all the three shutters in the experimental set-up of Fig. \ref{Fig3} are closed. The three peaks of the $6\text{P}_{3/2}(\text{F}=2,3,4)$ hyperfine levels are merged forming a broad transparency spectrum due to the residual Doppler broadening effect. When shutter 1 is open, dips corresponding to three hyperfine levels are induced inside the broad transparency peaks caused by VIPO at IR transition (see the blue trace marked with dots in Fig. \ref{Fig6}a). However, the dips appear very small due to the broad transparency background. The effect is removed when shutter 3 is open to subtract the broad transparency profile and the spectrum of the resolved hyperfine levels is shown by the green trace of Fig. \ref{Fig6}a. The linewidth of the resolved peaks are as follows: $\text{F}=4$ is 13.3~MHz, $\text{F}=3$ is 14.1~MHz and $\text{F}=2$ is 12.1~MHz. The power of the pump beams labeled c1, c2 and c3 used for optimal signal-to-noise ratio of the spectrum are 276.2~$\mu\text{W}$ (or peak intensity I=11.7~$\text{mW}/\text{cm}^2$), 5.02~mW (or peak intensity I=106.5~$\text{mW}/\text{cm}^2$) and 3.64~mW (or peak intensity I=77.2~$\text{mW}/\text{cm}^2$) respectively.
% Figure 6 %%%%%%%%%%%%%%%%%%%%%%%      
%%%%%%%%%%%%%%%%%%%%%%%%%%%%%%%%%%

Further line narrowing of the resolved peaks is achieved using the configuration shown in Fig. \ref{Fig1}c (i.e. VIPO at IR and VSS at blue transition). The energy configuration scheme in Fig. \ref{Fig1}c (i.e. VIPO at IR and VSS at blue transition) is implemented in the experimental set-up given in Fig. \ref{Fig3}, when shutter 1 and shutter 2 are open. Lower power of IR pump beam is used in this configuration since the induced dips by VIPO at IR are enhanced by VSS effect at blue transition. The transparency spectrum of this configuration is shown by the blue trace marked with dots in Fig. \ref{Fig6}b. The broad transparency background is removed when shutter 3 is open and the well resolved peaks of the $6\text{P}_{3/2}(\text{F}=2,3,4)$ hyperfine levels is shown by the green trace of Fig. \ref{Fig6}b. The linewidth of the resolved peaks are as follows: $\text{F}=4$ is 10.8~MHz, $\text{F}=3$ is 9.1~MHz and $\text{F}=2$ is 11.4~MHz. The power of the pump beams labeled c1, c2 and c3 used for optimal signal-to-noise ratio of the spectrum are 176.4~$\mu\text{W}$ (or peak intensity I=7.5~$\text{mW}/\text{cm}^2$), 6.01~mW (or peak intensity I=127.5~$\text{mW}/\text{cm}^2$) and 8.62~mW (or peak intensity I=182.9~$\text{mW}/\text{cm}^2$) respectively.
% Figure 7 %%%%%%%%%%%%%%%%%%%%%%%      
\begin{figure}
  \begin{center} 
  \includegraphics[width =1.0\linewidth]{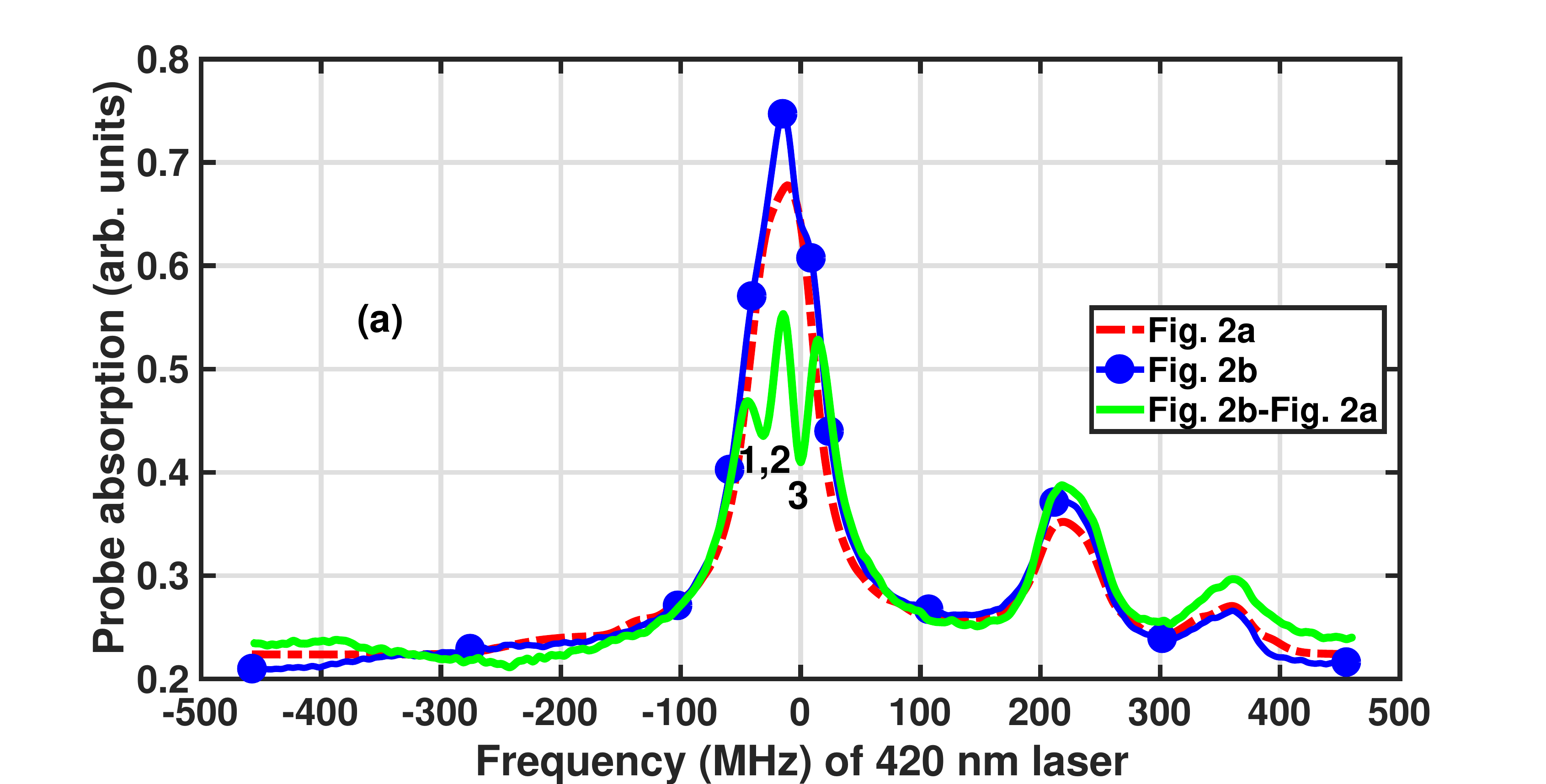}
  \includegraphics[width =1.0\linewidth]{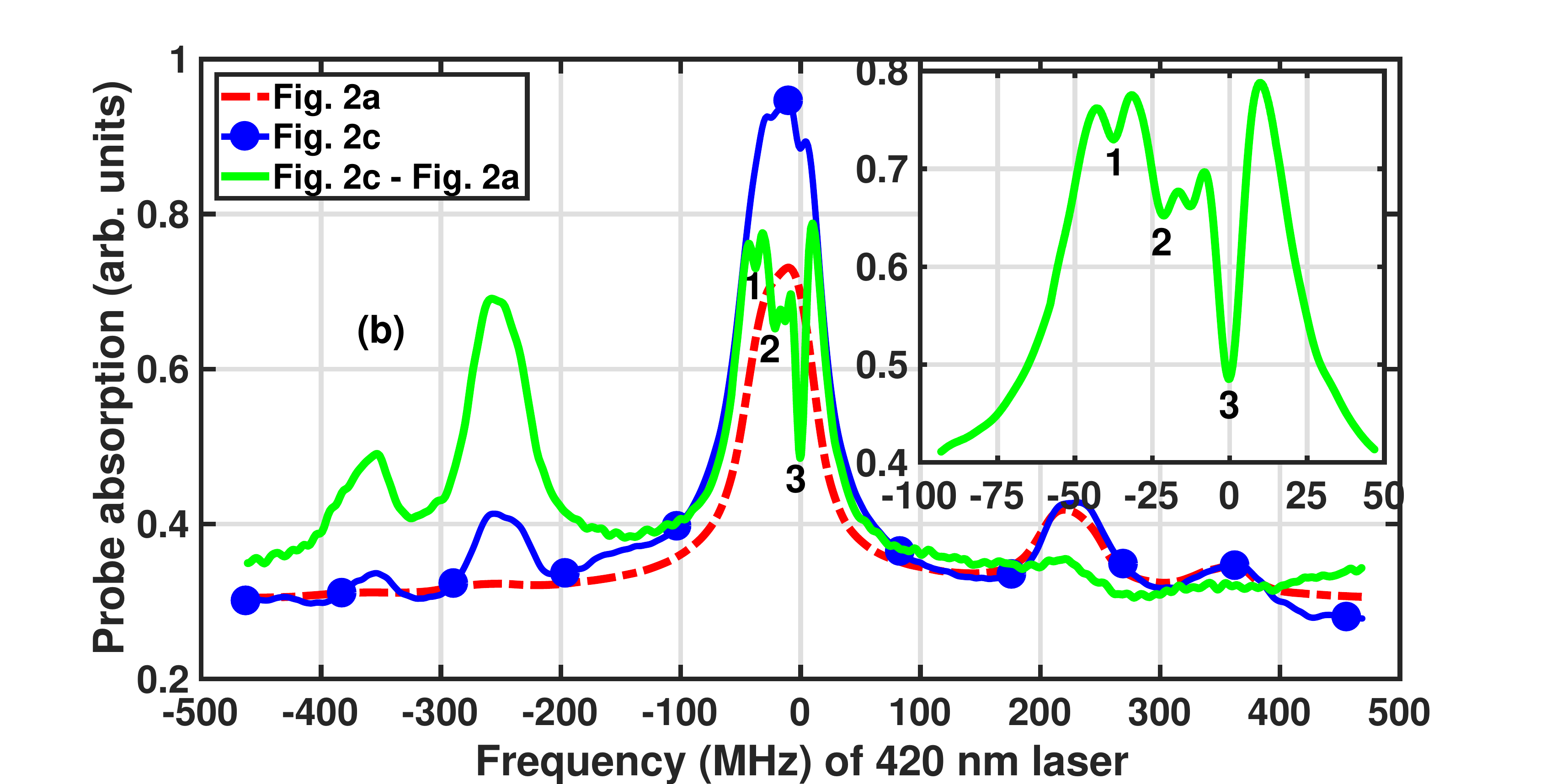}
    \caption{(Color online). The EA spectrum of the $6\text{P}_{3/2}$ hyperfine levels in $^{85}\text{Rb}$ under various configurations shown in Fig. \ref{Fig2}. The red dashed trace is for the optical pumping system (Fig. \ref{Fig2}a), the blue trace marked with dots in Fig. \ref{Fig7}a is for the optical pumping system with VIPO effect at IR transition (as shown in Fig. \ref{Fig2}b) while the blue trace marked with dots in Fig. \ref{Fig7}b is for the optical pumping system with VIPO effect at IR and VSS effect at blue transition (Fig. \ref{Fig2}c). The green trace is the final result after removing broad absorption background and it is magnified 3 time for visibility purpose.} 
    \label{Fig7}
   \end{center}
\end{figure}
%%%%%%%%%%%%%%%%%%%%%%%%%%%%%%%%%%

In the final result of the resolved peaks (see the green trace of Fig. \ref{Fig6}b), there are small peaks between the main peaks of $\text{F}=3$ and $\text{F}=4$ and between $\text{F}=2$ and $\text{F}=3$. These are not cross-over peaks (or real peaks), but the residue due to incomplete removal of the broad transparency background in the overlapped regions. The effect also occur for the optical pumping system when the broad absorption background is removed (see the green trace of Fig. \ref{Fig7}b the small peak between $\text{F}=2$ and $\text{F}=3$).

\subsubsection{Optical pumping system}

The EA spectrum of the optical pumping system is shown by the red dashed trace in Fig. \ref{Fig7}. This spectrum is obtained when all the three shutters in the experimental set-up of Fig. \ref{Fig3} are closed. The absorption peaks corresponding to the $6\text{P}_{3/2}(\text{F}=1,2,3)$ hyperfine levels are completely merged. The levels $6\text{P}_{3/2}(\text{F}=2,3)$ are detected by the probe via both the direct decay and indirect decay channels \cite{RMS12} while level $6\text{P}_{3/2}(\text{F}=1)$ is detected via the indirect decay channels to $5\text{S}_{1/2}(\text{F}=3)$ only. When shutter 1 is open, dips corresponding to the hyperfine levels are induced inside the broad EA peaks due to VIPO at IR transition (see the blue trace marked with dots in Fig. \ref{Fig7}a). The dips appear small due to broad EA background caused by the residual Doppler broadening effect. The broad EA background is removed when shutter 3 is open and the dips corresponding to the hyperfine levels $6\text{P}_{3/2}(\text{F}=1,2)$ are still not resolved while the $6\text{P}_{3/2}(\text{F}=3)$ peak is resolved (see the green trace of Fig. \ref{Fig7}a). The linewidth of the resolved peak is $\text{F}=3$ is 13.9~MHz. The power of the pump beams labeled c1, c2 and c3 used for optimal signal-to-noise ratio of the spectrum are 806.2~$\mu\text{W}$ (or peak intensity I=34.2~$\text{mW}/\text{cm}^2$), 5.01~mW (or peak intensity I=106.3~$\text{mW}/\text{cm}^2$) and 1.84~mW (or peak intensity I=39.0~$\text{mW}/\text{cm}^2$) respectively.

The peaks corresponding to the $6\text{P}_{3/2}(\text{F}=1,2,3)$ hyperfine levels, can be completely resolved using the configuration shown in Fig. \ref{Fig2}c i.e. VIPO at IR and VSS at blue transition. This configuration is implemented when shutter 1 and shutter 2 are open in the experimental set-up of Fig. \ref{Fig3}. The broad EA spectrum is removed when shutter 3 is open and the green trace of Fig. \ref{Fig7}b shows well resolved peaks of the $6\text{P}_{3/2}(\text{F}=1,2,3)$ hyperfine levels. Note, the frequency scaling of the spectra in Fig. \ref{Fig7} is assigned using the peak locations of $\text{F}=2$ and $\text{F}=3$ after the complete resolution of all the three peaks of $6\text{P}_{3/2}(\text{F}=1,2,3)$ hyperfine levels. The linewidth of the resolved peaks are as follows: $\text{F}=3$ is 9.8~MHz, $\text{F}=2$ is 10.1~MHz and $\text{F}=1$ is 7.2~MHz. The power of the pump beams labeled c1, c2 and c3 used for optimal signal-to-noise ratio of the spectrum are 276.3~$\mu\text{W}$ (or peak intensity I=11.7~$\text{mW}/\text{cm}^2$), 5.02~mW (or peak intensity I=106.7~$\text{mW}/\text{cm}^2$) and 15.19~mW (or peak intensity I=322.3~$\text{mW}/\text{cm}^2$) respectively.

Besides the main peaks due to near zero-velocity group atoms in Fig. \ref{Fig7}a, the extra peaks (or cross-over peaks) formed outside the main spectrum are caused by atoms moving with velocities of 94 ms$^{-1}$ and 143 ms$^{-1}$ respectively. Atoms moving with velocities of 94 ms$^{-1}$ and 143ms$^{-1}$ along the propagation direction of the IR probe, will see the probe laser to be on resonance with the $5\text{S}_{1/2}(\text{F}=3)\rightarrow5\text{P}_{3/2}(\text{F}=3)$ and $5\text{S}_{1/2}(\text{F}=3)\rightarrow5\text{P}_{3/2}(\text{F}=2)$ transitions respectively. The corresponding extra peaks location will be at 224 MHz and 342 MHz from the main peaks. In Fig. \ref{Fig7}b, the counter-propagating blue laser beams will form extra peaks on both the left and right side of the main peaks. Ideally the extra peak on the right side of the green spectrum should vanish, but it is still visible due to incomplete subtraction.

\subsection{Resolving $6\text{P}_{3/2}$ hyperfine levels in $^{87}\text{Rb}$}

\subsubsection{V-type open system}

The $6\text{P}_{3/2}$ hyperfine levels of $^{87}\text{Rb}$ were also resolved using similar configurations shown in Fig. \ref{Fig1} and \ref{Fig2}. The results of VIPO at IR plus VSS at blue transition configuration both in the case of a V-type open system and optical pumping system are reported here. In this configuration, the probe and the counter-propagating pump lasers at $780~\text{nm}$ are locked to resonance on $5\text{S}_{1/2}(\text{F}=2)\leftrightarrow5\text{P}_{3/2}(\text{F}=3)$ transition. The $420~\text{nm}$ pump laser scans across the $6\text{P}_{3/2}$ hyperfine levels on $5\text{S}_{1/2}(\text{F}=2)\leftrightarrow6\text{P}_{3/2}$ weak transition in the case of a V-type system and on $5\text{S}_{1/2}(\text{F}=1)\leftrightarrow6\text{P}_{3/2}$ weak transition in  the case of optical pumping system.
% Figure 8 %%%%%%%%%%%%%%%%%%%%%%%      
\begin{figure}
   \begin{center}    
    \includegraphics[width =1.0\linewidth]{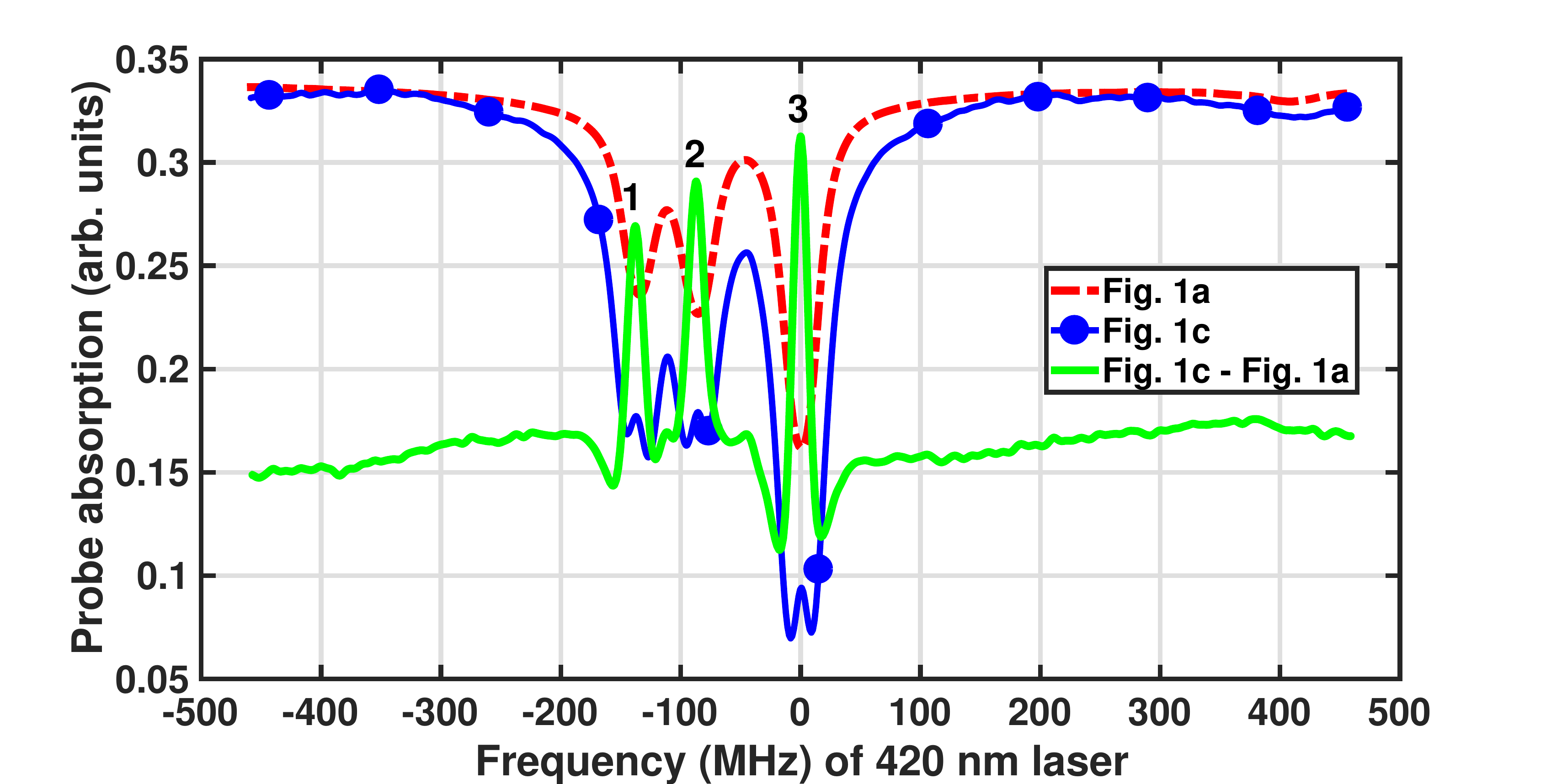}
      \caption{(Color online). The transparency spectrum of the $6\text{P}_{3/2}$ hyperfine levels in $^{87}\text{Rb}$ recorded for similar configurations of $^{85}\text{Rb}$ shown in Fig. \ref{Fig1}. The red dashed trace is for the V-type open system (Fig. \ref{Fig1}a), the blue trace marked with dots is for V-type open system with VIPO effect at IR and VSS effect at blue transition (Fig. \ref{Fig1}c) and the green trace is the final result after removing the broad transparency background and it is magnified 3 times for visibility purpose.}
      \label{Fig8}
   \end{center}
\end{figure}
%%%%%%%%%%%%%%%%%%%%%%%%%%%%%%%%%% 

The transparency spectrum of the configuration given in Fig. \ref{Fig1}a for $^{87}\text{Rb}$, is shown by the red dashed trace of Fig. \ref{Fig8} when all the three shutters in the experimental set-up of Fig. \ref{Fig3} are closed. The peaks of the $6\text{P}_{3/2}(\text{F}=2,3)$ hyperfine levels are well resolved but the peaks of $6\text{P}_{3/2}(\text{F}=1,2)$ are partially resolved due to the residual Doppler broadening effect. When shutters 1 and 2 are open, the dips induced by VIPO at IR and VSS at blue transition inside the broad transparency peaks corresponds to the three hyperfine levels of the $6\text{P}_{3/2}(\text{F}=1,2,3)$ state (see the blue trace marked with dots in Fig. \ref{Fig8}). The residual Doppler broadening effect is removed when shutter 3 is open and the spectrum of the resolved hyperfine levels is shown by the green trace of Fig. \ref{Fig8}. The linewidth of the resolved peaks are as follows: $\text{F}=3$ is 14.4~MHz, $\text{F}=2$ is 15.7~MHz and $\text{F}=1$ is 15.8~MHz. The power of the pump beams labeled c1, c2 and c3 used for optimal signal-to-noise ratio of the spectrum are $302.6~\mu\text{W}$ (or peak intensity I=12.8~$\text{mW}/\text{cm}^2$), 4.82~mW (or peak intensity I=102.3~$\text{mW}/\text{cm}^2$) and 13.2~mW (or peak intensity I=280.1~$\text{mW}/\text{cm}^2$) respectively.

\subsubsection{Optical pumping system}
\begin{figure}
   \begin{center}    
    \includegraphics[width =1.0\linewidth]{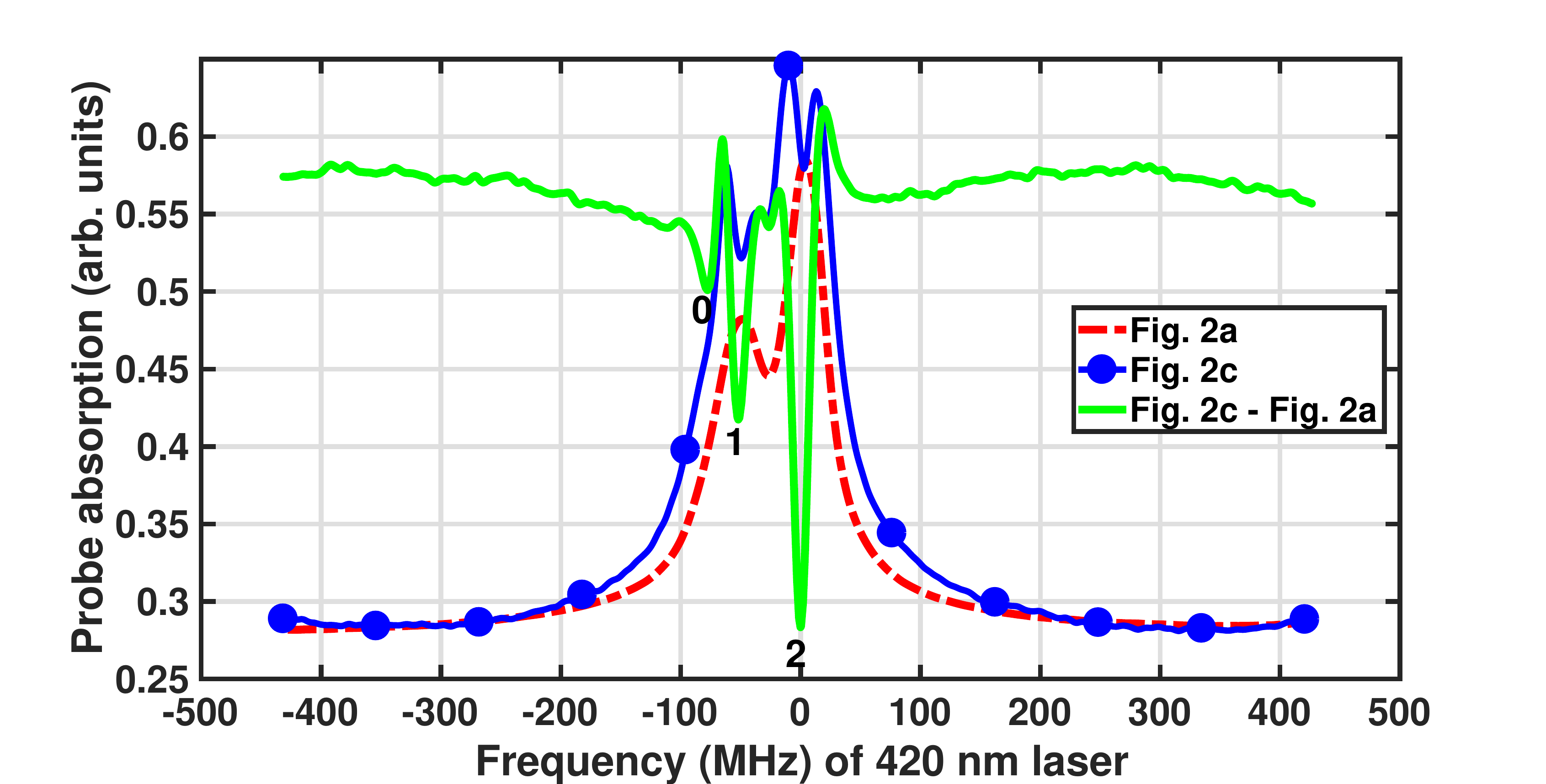}
      \caption{(Color online). The EA spectrum of the $6\text{P}_{3/2}$ hyperfine levels in $^{87}\text{Rb}$ recorded for similar configurations of $^{85}\text{Rb}$ shown in Fig. \ref{Fig2}. The red dashed trace is for the optical pumping system (Fig. \ref{Fig2}a), the blue trace marked with dots is for the optical pumping system with VIPO effect at IR and VSS effect at blue transition (Fig. \ref{Fig2}c) and the green trace is the final result after removing the broad absorption background and it is magnified 3 times for visibility purpose}
      \label{Fig9}
   \end{center}
\end{figure}

The EA spectrum of the optical pumping system is shown by the red dashed trace in Fig. \ref{Fig9} when all the three shutters in the experimental set-up of Fig. \ref{Fig3} are closed. The absorption peaks corresponds to the $6\text{P}_{3/2}(\text{F}=0,1,2)$ hyperfine levels in $^{87}\text{Rb}$. The peaks for $6\text{P}_{3/2}(\text{F}=0,1)$ are completely merged while the peaks for $6\text{P}_{3/2}(\text{F}=1,2)$ are partially merged. The levels $6\text{P}_{3/2}(\text{F}=1,2)$ are detected by the probe via both the direct decay and indirect decay channels \cite{RMS12} while level $6\text{P}_{3/2}(\text{F}=0)$ is detected via the indirect decay channels to $5\text{S}_{1/2}(\text{F}=2)$ only. When shutters 1 and 2 are open, dips corresponding to the hyperfine levels are induced inside the broad EA peaks due to VIPO at IR and VSS at blue transition (see the blue trace marked with dots in Fig. \ref{Fig9}). The dips appear small due to broad EA background caused by the residual Doppler broadening effect. The broad EA background is removed when shutter 3 is open and the dips corresponding to the hyperfine levels $6\text{P}_{3/2}(\text{F}=0,1,2)$ are resolved (see the green trace of Fig. \ref{Fig9}). The linewidth of the resolved peaks are as follows: $\text{F}=2$ is $16.4~\text{MHz}$, $\text{F}=1$ is $13.1~\text{MHz}$ and $\text{F}=0$ is $12.3~\text{MHz}$. The power of the pump beams labeled c1, c2 and c3 used for optimal signal-to-noise ratio of the spectrum are $823.6~\mu\text{W}$ (or peak intensity I=35.0~$\text{mW}/\text{cm}^2$), 4.82~mW (or peak intensity I=102.3~$\text{mW}/\text{cm}^2$) and 15.3~mW (or peak intensity I=325.5~$\text{mW}/\text{cm}^2$) respectively.
%% Figure 9 %%%%%%%%%%%%%%%%%%%%%%%      
%%%%%%%%%%%%%%%%%%%%%%%%%%%%%%%%%% 

\subsubsection{Power broadening effect}

\begin{figure}
   \begin{center} 
   \includegraphics[width =1.0\linewidth]{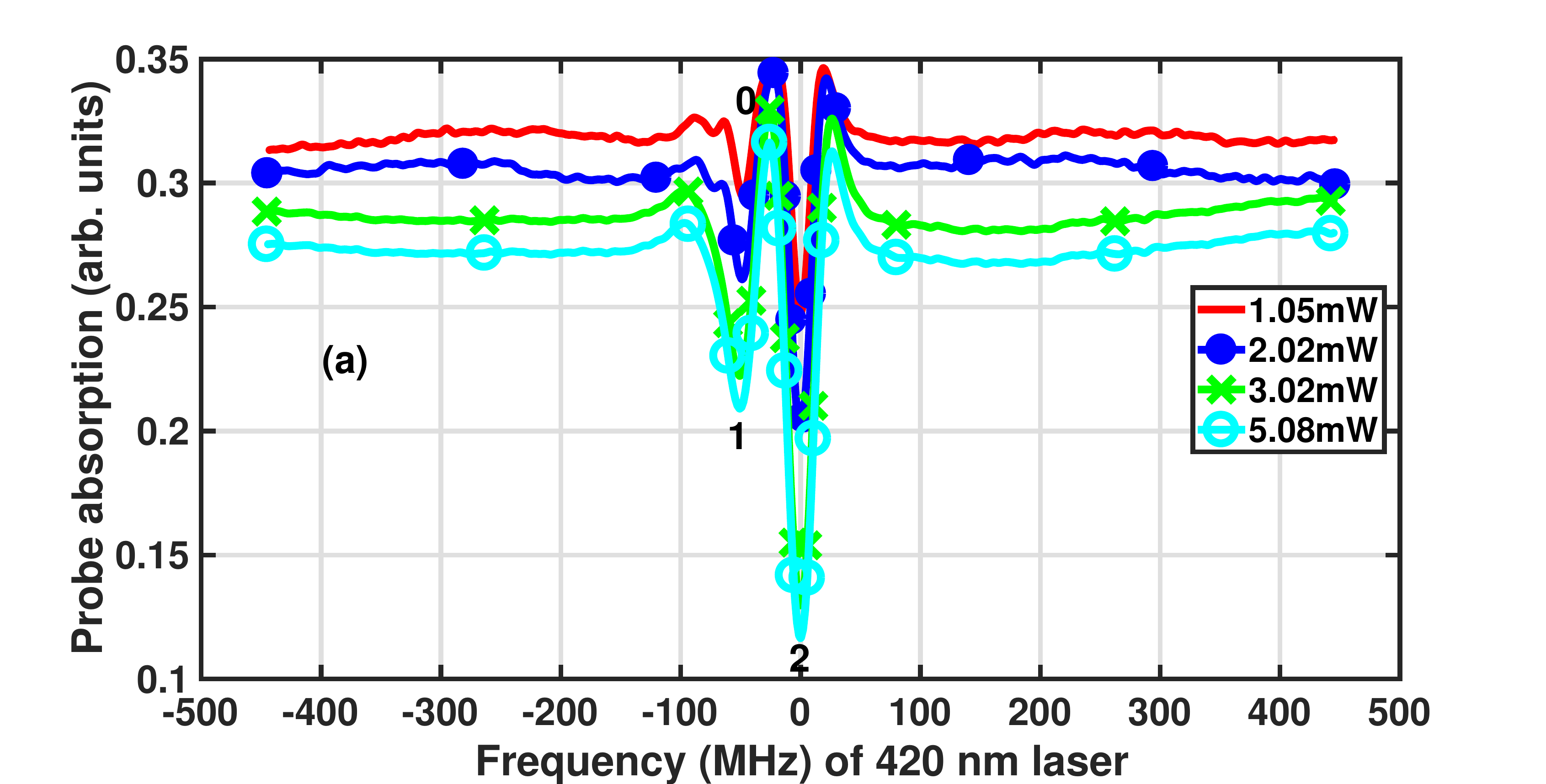}   
    \includegraphics[width =1.0\linewidth]{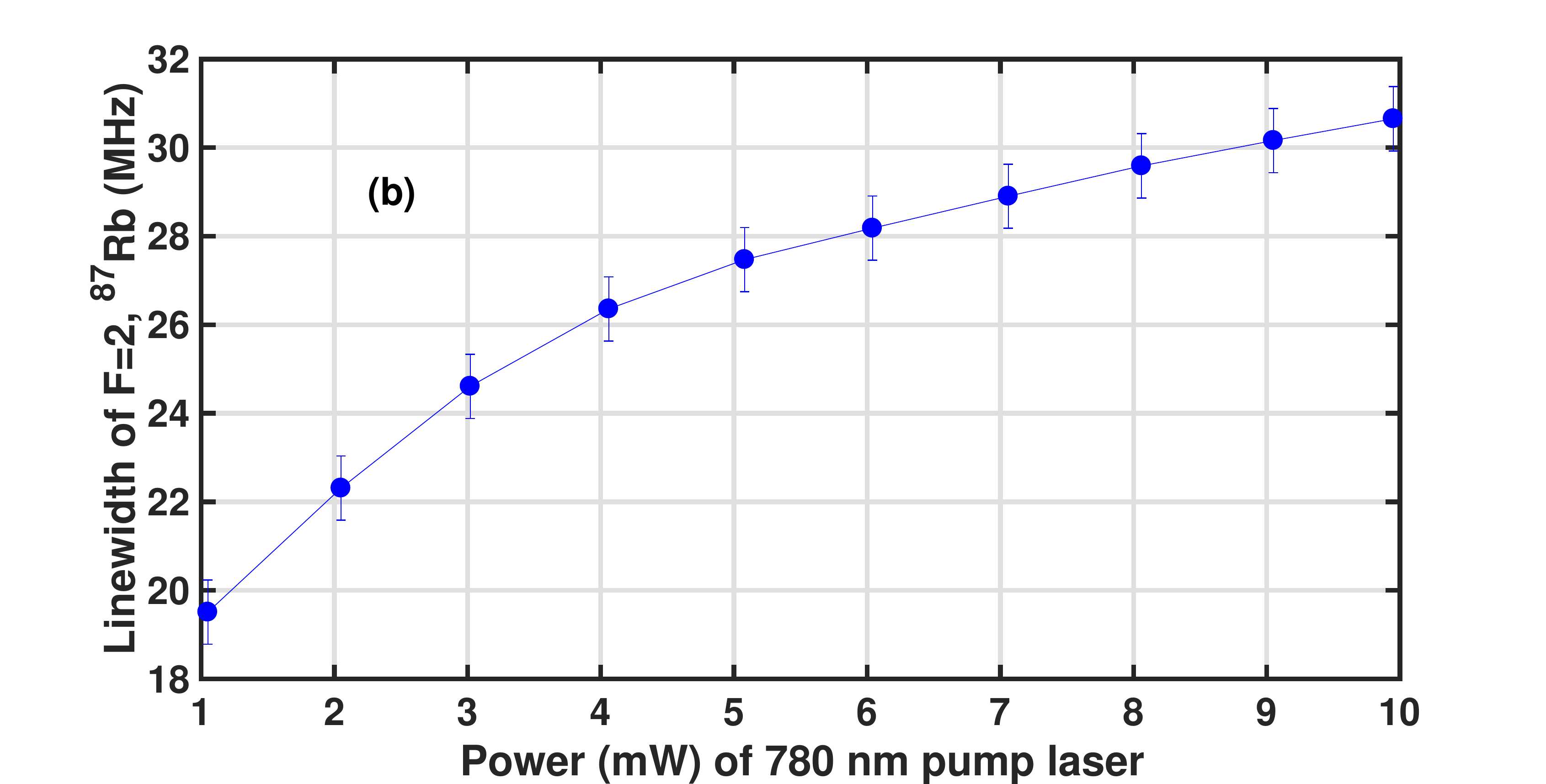}
      \caption{(Color online). (a) The result of the final spectra of the VIPO dips of the $6\text{P}_{3/2}$ hyperfine levels in $^{87}\text{Rb}$ recorded for various powers of IR pump laser after removing the broad absorption background. (b) The variation of the linewidth of the resolved peak $\text{F}=2$ with the power of IR pump.}
      \label{Fig10}
   \end{center}
\end{figure}

The contribution of the IR pump power broadening effect to the final result (i.e. the resolved spectrum of the $6\text{P}_{3/2}$ state), is illustrated in Fig. \ref{Fig10}a. The configuration used here is given in Fig. \ref{Fig2}b (i.e. VIPO at IR transition) for the case of $^{87}\text{Rb}$. The power of the blue pump laser beams is fixed (i.e. c2 is 4.26 mW and C3 is 3.27 mW) as the the power of IR pump is changed. At 1 mW of the IR pump, all the three peaks corresponding to the $6\text{P}_{3/2}(\text{F}=0,1,2)$ hyperfine levels are well resolved (see the red trace of Fig. \ref{Fig10}a). However, as the IR pump power is increased to 5 mW, the peaks corresponding to the $6\text{P}_{3/2}(\text{F}=0,1)$ are completely merged as shown by the cyan trace marked by circles in Fig. \ref{Fig10}a. High intensity of the IR pump broadens the VIPO dips and limits the resolution of the closely spaced hyperfine levels of $\text{F}=0$ and $\text{F}=1$ which are 23.739~MHz apart \cite{GKK20}. The frequency scaling of the spectra in Fig. \ref{Fig10}a is assigned using the resolved peak locations of $\text{F}=1$ and $\text{F}=2$ of the red trace. The variation of the linewidth of the resolved peak corresponding to $\text{F}=2$ with the IR pump power is shown in Fig. \ref{Fig10}b. 

%% Figure 10 %%%%%%%%%%%%%%%%%%%%%%%      
%%%%%%%%%%%%%%%%%%%%%%%%%%%%%%%%%% 

\section{Conclusions}\label{Conc}

In conclusion we have presented a detailed experimental technique to eliminate the residual (or partial) Doppler broadening in a Doppler mismatched double resonance spectroscopy for a transparency spectrum (or enhanced absorption spectrum). The residual two-photon Doppler broadening is removed using the VIPO at IR transition, VSS at blue transition and the combination of the two effects followed by the subtraction of the broad transparency background or EA background. The technique has been used to resolve the closely spaced hyperfine levels of weak transitions for a Doppler mismatched double resonance at $780~\text{nm}$ and $420~\text{nm}$ in Rb at room temperature.

\section*{Acknowledgement}
%\ack{Acknowledgment}
E.O.N. would like to acknowledge Indian Council for Cultural Relations (ICCR) for the PhD scholarship. K.P. would like to acknowledge the funding from SERB of grant No. ECR/2017/000781.

\appendix
\appendixpage
\addappheadtotoc
\counterwithin{figure}{section}

\section{Transparency for V-type open system}
\label{AppendixA}

\subsection{V-type open system}\label{AppendA1}
\begin{figure}
   \begin{center}    
    \includegraphics[width =1.0\linewidth]{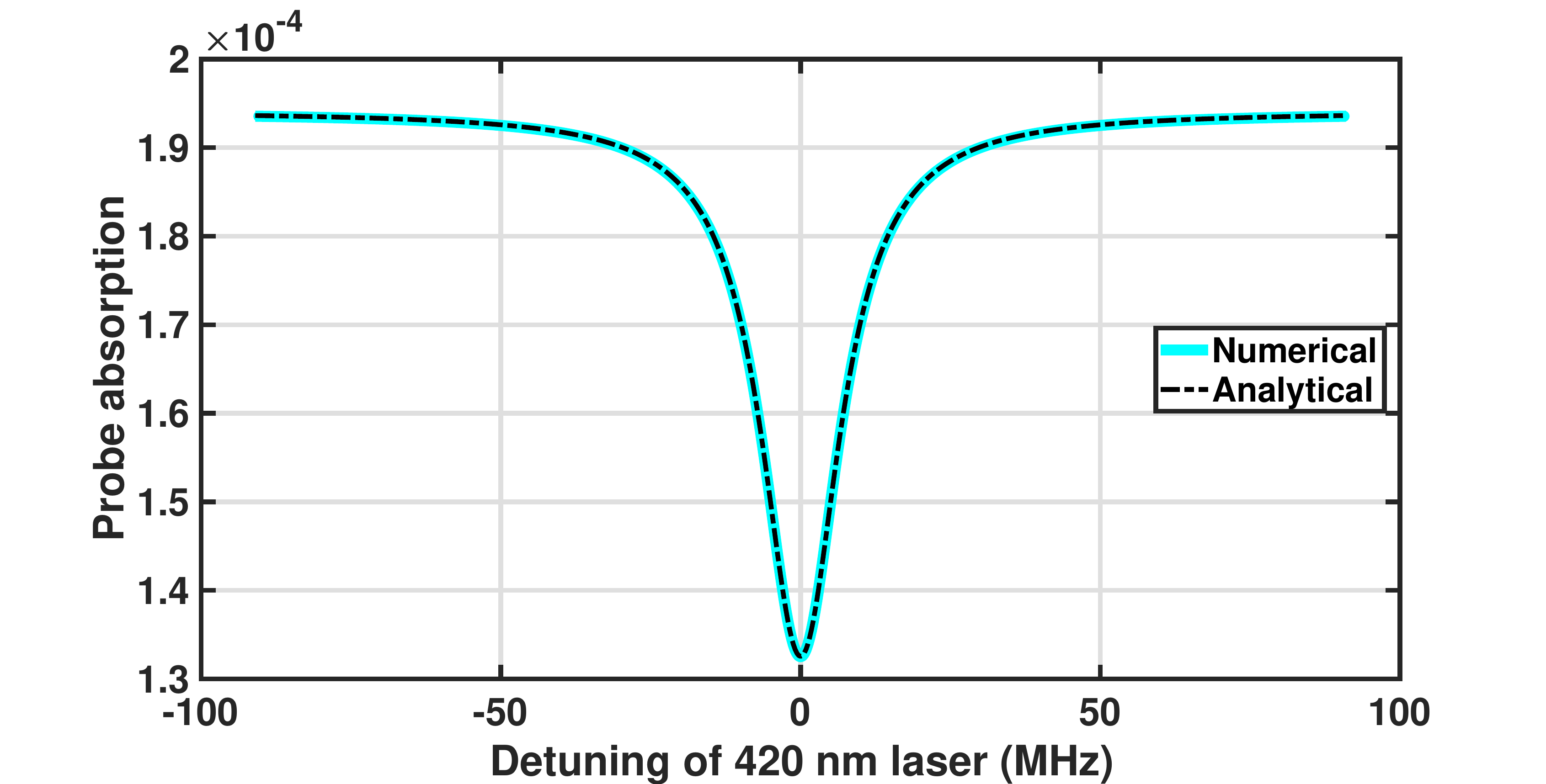}
      \caption{(Color online). Comparison of the full numerical solution of the density matrix in a V-type open system with the the analytical solution given in Eq. \ref{eqA3}. $\Omega_{c2}=\sqrt{1.5}\Gamma_{3}$, $\Gamma_{2}=2\pi\times6.065~\text{MHz},~\Gamma_{3}=2\pi\times1.32~\text{MHz}$.}
      \label{FigA1}
   \end{center}
\end{figure}

The Hamiltonian $\text{H}$ of a V-type open system shown in Fig. \ref{Fig1}a under electric-dipole and rotating-wave approximation and in the interaction picture is given as follows,
\begin{align}
\label{eqA1}
\text{H}=&\frac{\hbar}{2}\big\{\Omega_{p}\ket{1}\bra{2}+\Omega_{c2}\ket{1}\bra{3}-\Delta_{p}\ket{2}\bra{2}-\Delta_{c2}\ket{3}\bra{3}+h.c.\big\}
\end{align}
where the levels are $5\text{S}_{1/2}(\text{F}=3)=\ket{1}$, $5\text{P}_{3/2}(\text{F}=4)=\ket{2}$, $6\text{P}_{3/2}(\text{F}=2)=\ket{3}$ and $5\text{S}_{1/2}(\text{F}=2)=\ket{4}$. The equations of motion of the density matrix are obtained using Eq. \ref{eq2} and \ref{eqA1} and are given as follows,
\begin{align}
\label{eqA2}
\dot{\rho}_{12}=&\frac{i\Omega_{p}}{2}{(\rho_{11}-\rho_{22})}-\frac{i\Omega_{c2}}{2}\rho_{32}-\gamma_{12}\rho_{12}\nonumber\\
\dot{\rho}_{13}=&\frac{i\Omega_{c2}}{2}{(\rho_{11}-\rho_{33})}-\gamma_{13}\rho_{13}-\frac{i\Omega_{p}}{2}{\rho_{23}}\nonumber\\
\dot{\rho}_{14}=&-\frac{i\Omega_{c2}}{2}{\rho_{34}}-\gamma_{14}\rho_{14}-\frac{i\Omega_{p}}{2}{\rho_{24}}\nonumber\\
\dot{\rho}_{22}=&\frac{i\Omega_\text{p}}{2}{\rho_{21}}-\frac{i\Omega_{p}^{\ast}}{2}{\rho_{12}}-\Gamma_{2}{\rho_{22}}\nonumber\\
\dot{\rho}_{23}=&-\frac{i\Omega_{p}^{\ast}}{2}{\rho_{13}}+\frac{i\Omega_{c2}}{2}{\rho_{21}}-\gamma_{23}\rho_{23}\nonumber\\
\dot{\rho}_{24}=&-\frac{i\Omega_{p}^{\ast}}{2}{\rho_{14}}-\gamma_{24}\rho_{24}\nonumber\\
\dot{\rho}_{33}=&-\frac{i\Omega_{c2}^{\ast}}{2}{\rho_{13}}+\frac{i\Omega_{c2}}{2}{\rho_{31}}-\Gamma_{3}{\rho_{33}}\\
\dot{\rho}_{34}=&-\frac{i\Omega_{c2}^{\ast}}{2}{\rho_{14}}-\gamma_{34}{\rho_{34}}\nonumber\\
\dot{\rho}_{44}=&\Gamma_{34}{\rho_{33}}+\Pi_{g}{(\rho_{11}-\rho_{44})}\nonumber
\end{align}
where, $\gamma_{12}=i\Delta_{p}+\gamma^{dec}_{12}$, $\gamma_{13}=i\Delta_{c2}+\gamma^{dec}_{13}$, $\gamma_{14}=\gamma^{dec}_{14}$, $\gamma_{23}=i(\Delta_{c2}-\Delta_{p})+\gamma^{dec}_{23}$, $\gamma_{24}=-i\Delta_{p}+\gamma^{dec}_{24}$, $\gamma_{34}=-i\Delta_{c2}+\gamma^{dec}_{34}$, $\Gamma_{1}=\Gamma_{4}=\Pi_{g}$, $\Gamma_{3}=\Gamma_{31}+\Gamma_{34}$, and $\gamma^{dec}_{ij}={\frac{1}{2}}(\Gamma_{i}+\Gamma_{j})$, $\Gamma_{i}$ is the decay rate of the $i^{\text{th}}$ level, $\Gamma_{31}$ and $\Gamma_{34}$ are the decay rates of level 3 to level 1 and level 4 respectively. The remaining density matrix equations are obtained using population conservation law $\sum_{j=1}^{4}\rho_{jj}=1$ and the complex conjugate $\dot{\rho}_{ji}=\dot{\rho}^{\ast}_{ij}$. In the steady state condition ($\dot{\rho}_{ij}= 0$ for all $i$ and $j$), the imaginary part of $\rho_{12}$ corresponds to the absorption of the probe laser and in the weak probe approximation it is given as follows, 
\begin{widetext}
\begin{align}
\label{eqA3}
\rho_{12}=\frac{i \Pi_{g} \Omega_{p} (\Gamma_{3} ((\Gamma_{3}+\Pi_{g})^2+4 \Delta_{c}^2)+\frac{\Omega_{c}^2 (\Gamma_{2} (\Gamma_{3}+\Pi_{g})-2 i \Delta_{c} (2 \Gamma_{3}+\Pi_{g})+2 i \Delta_{p} (\Gamma_{3}+\Pi_{g}))}{\Gamma_{2}+\Gamma_{3}-2 i (\Delta_{c}-\Delta_{p})})}{(\Omega_{c}^2 (\Gamma_{3}+\Pi_{g}) (\Gamma_{34}+3 \Pi_{g})+2 \Gamma_{3} \Pi_{g} ((\Gamma_{3}+\Pi_{g})^2+4 \Delta_{c}^2)) (\frac{\Omega_{c}^2}{\Gamma_{2}+\Gamma_{3}-2 i \Delta_{c}+2 i \Delta_{p}}+\Gamma_{2}+\Pi_{g}+2 i \Delta_{p})}
\end{align}
\end{widetext}
\normalsize{}The solution of a V-type open system given in Eq. \ref{eqA3} is graphically represented in Fig. \ref{FigA1} and is well matched with the numerical simulation of the full density matrix given in Eq. \ref{eqA2}.
% Figure A1 %%%%%%%%%%%%%%%%%%%%%%%      
%%%%%%%%%%%%%%%%%%%%%%%%%%%%%%%%%% 

\subsection{VIPO at IR transition for V-type open system}
\begin{figure}
   \begin{center}    
    \includegraphics[width =1.0\linewidth]{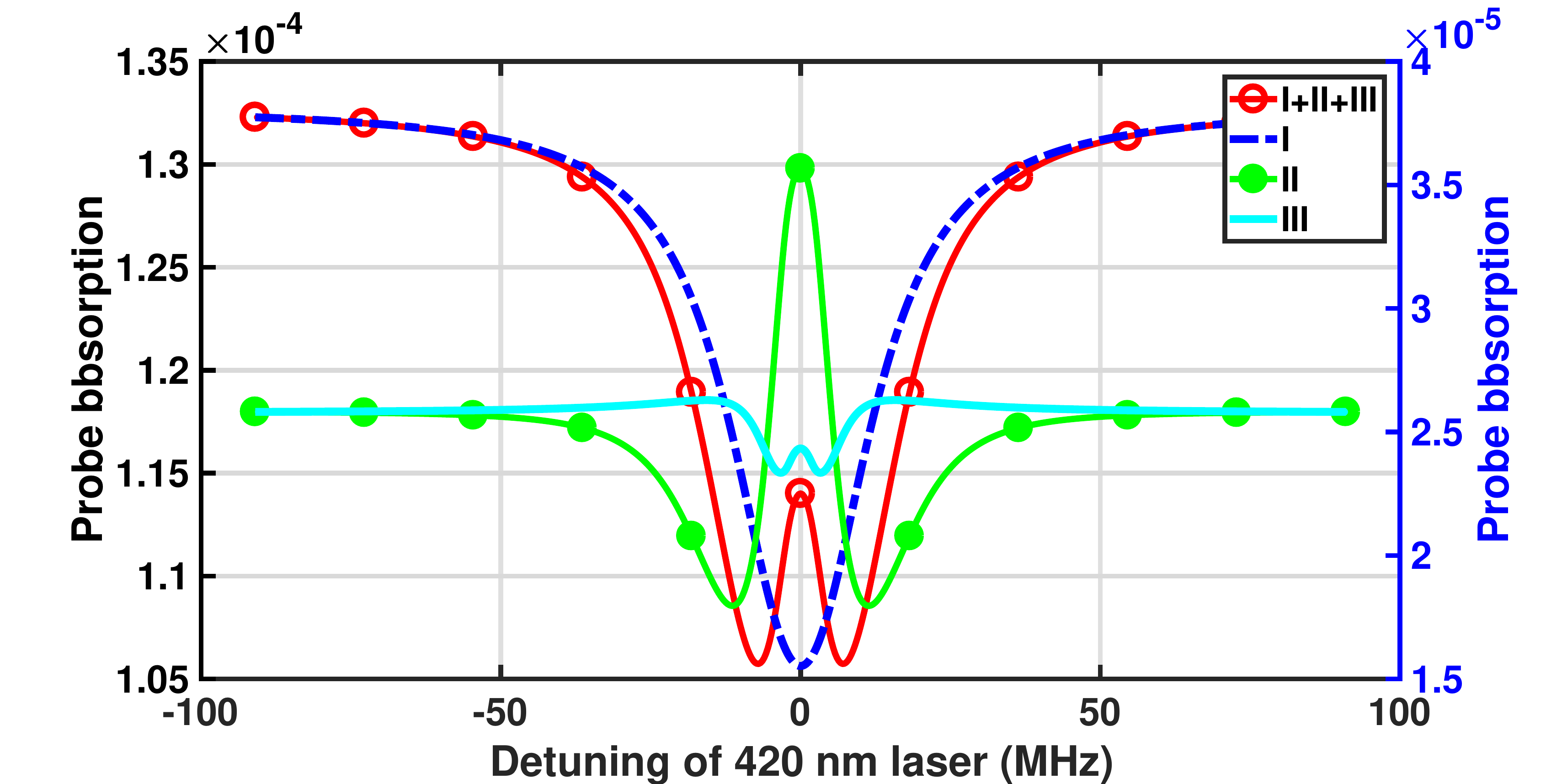}
      \caption{(Color online). The graphical representation of the individual terms I, II and III given in Eq. \ref{eq4}. $\Omega_{c1}=\Gamma_{2}$, $\Omega_{c2}=\sqrt{1.5}\Gamma_{3}$, $\Gamma_{2}=2\pi\times6.065~\text{MHz},~\Gamma_{3}=2\pi\times1.32~\text{MHz}$. The vertical axis of the red trace marked by circles and the blue dashed trace are on the left and the green trace marked with dots and the cyan trace are on the right.}
      \label{FigA2}
   \end{center}
\end{figure}

The Hamiltonian of the VIPO at IR transition for a V-type open system shown in Fig. \ref{Fig1}b is given in Eq. \ref{eq1}. The following set of equations of motion are obtained by substitution of Eq. \ref{eq1} into Eq. \ref{eq2}.
\small{}
\begin{align}
\label{eqA4}
\dot{\rho}_{12}=&\frac{i}{2}{(\Omega_\text{c1}+\Omega_\text{p}e^{i\delta_{1}{t}})(\rho_{11}-\rho_{22})}-\frac{i\Omega_\text{c2}}{2}\rho_{32}-\gamma_{12}\rho_{12}\nonumber\\
\dot{\rho}_{13}=&\frac{i\Omega_\text{c2}}{2}{(\rho_{11}-\rho_{33})}-\gamma_{13}\rho_{13}-\frac{i}{2}{(\Omega_\text{c1}+\Omega_\text{p}e^{i\delta_{1}{t}})\rho_{23}}\nonumber\\
\dot{\rho}_{14}=&-\frac{i\Omega_\text{c2}}{2}{\rho_{34}}-\gamma_{14}\rho_{14}-\frac{i}{2}{(\Omega_\text{c1}+\Omega_\text{p}e^{i\delta_{1}{t}})\rho_{24}}\nonumber\\
\dot{\rho}_{22}=&\frac{i}{2}{(\Omega_\text{c1}+\Omega_\text{p}e^{i\delta_{1}{t}})\rho_{21}}-\frac{i}{2}{(\Omega_\text{c1}^{\ast}+\Omega_\text{p}^{\ast}e^{-i\delta_{1}{t}})\rho_{12}}-\Gamma_{2}{\rho_{22}}\nonumber\\
\dot{\rho}_{23}=&-\frac{i}{2}{(\Omega_\text{c1}^{\ast}+\Omega_\text{p}^{\ast}e^{-i\delta_{1}{t}})\rho_{13}}+\frac{i\Omega_\text{c2}}{2}{\rho_{21}}-\gamma_{23}\rho_{23}\nonumber\\
\dot{\rho}_{24}=&-\frac{i}{2}{(\Omega_\text{c1}^{\ast}+\Omega_\text{p}^{\ast}e^{-i\delta_{1}{t}})\rho_{14}}-\gamma_{24}\rho_{24}\nonumber\\
\dot{\rho}_{33}=&-\frac{i\Omega_\text{c2}^{\ast}}{2}{\rho_{13}}+\frac{i\Omega_\text{c2}}{2}{\rho_{31}}-\Gamma_{3}{\rho_{33}}\\
\dot{\rho}_{34}=&-\frac{i\Omega_\text{c2}^{\ast}}{2}{\rho_{14}}-\gamma_{34}{\rho_{34}}\nonumber\\
\dot{\rho}_{44}=&\Gamma_{34}{\rho_{33}}+\Pi_{g}{(\rho_{11}-\rho_{44})}\nonumber
\end{align}
\normalsize{}where, $\gamma_{12}=i\Delta_\text{c1}+\gamma^{dec}_{12}$, $\gamma_{13}=i\Delta_\text{c2}+\gamma^{dec}_{13}$, $\gamma_{14}=\gamma^{dec}_{14}$, $\gamma_{23}=i(\Delta_\text{c2}-\Delta_\text{c1})+\gamma^{dec}_{23}$, $\gamma_{24}=-i\Delta_\text{c1}+\gamma^{dec}_{24}$, $\gamma_{34}=-i\Delta_\text{c2}+\gamma^{dec}_{34}$. The steady state solution of the equations of motion given in Eq. \ref{eqA4} is given in Eq. \ref{eq4} and the individual contribution of the terms I, II and III is illustrated in Fig. \ref{FigA2}. 
% Figure A2 %%%%%%%%%%%%%%%%%%%%%%%      
%%%%%%%%%%%%%%%%%%%%%%%%%%%%%%%%%%  

\subsection{VIPO at IR and VSS at blue transition for V-type open system}
The Hamiltonian of the VIPO at IR transition for a V-type open system considered in Fig. \ref{Fig1}b is given in Eq. \ref{eq6}. The following set of equations of motion of the density matrix are similarly obtained by substitution of Eq. \ref{eq6} into Eq. \ref{eq2}.
\small{}
\begin{align}
\label{eqA5}
\dot{\rho}_{12}=&\frac{i}{2}{(\Omega_\text{c1}+\Omega_\text{p}e^{i\delta_{1}{t}})(\rho_{11}-\rho_{22})}-\gamma_{12}\rho_{12}\nonumber\\
&-\frac{i\Omega_\text{c2}}{2}{(1+e^{i\delta_{2}{t}})}\rho_{32}\\
\dot{\rho}_{13}=&\frac{i\Omega_\text{c2}}{2}{(1+e^{i\delta_{2}{t}})}{(\rho_{11}-\rho_{33})}-\gamma_{13}\rho_{13}-\frac{i}{2}{(\Omega_\text{c1}+\Omega_\text{p}e^{i\delta_{1}{t}})\rho_{23}}\nonumber\\
\dot{\rho}_{14}=&-\frac{i\Omega_\text{c2}}{2}{(1+e^{i\delta_{2}{t}})}{\rho_{34}}-\gamma_{14}\rho_{14}-\frac{i}{2}{(\Omega_\text{c1}+\Omega_\text{p}e^{i\delta_{1}{t}})\rho_{24}}\nonumber\\
\dot{\rho}_{22}=&\frac{i}{2}{(\Omega_\text{c1}+\Omega_\text{p}e^{i\delta_{1}{t}})\rho_{21}}-\frac{i}{2}{(\Omega_\text{c1}^{\ast}+\Omega_\text{p}^{\ast}e^{-i\delta_{1}{t}})\rho_{12}}
-\Gamma_{2}{\rho_{22}}\nonumber\\
\dot{\rho}_{23}=&-\frac{i}{2}{(\Omega_\text{c1}^{\ast}+\Omega_\text{p}^{\ast}e^{-i\delta_{1}{t}})\rho_{13}}+\frac{i\Omega_\text{c2}}{2}{(1+e^{i\delta_{2}{t}})}{\rho_{21}}-\gamma_{23}\rho_{23}\nonumber\\
\dot{\rho}_{24}=&-\frac{i}{2}{(\Omega_\text{c1}^{\ast}+\Omega_\text{p}^{\ast}e^{-i\delta_{1}{t}})\rho_{14}}-\gamma_{24}\rho_{24}\nonumber\\
\dot{\rho}_{33}=&-\frac{i\Omega_\text{c2}^{\ast}}{2}{(1+e^{-i\delta_{2}{t}})}{\rho_{13}}+\frac{i\Omega_\text{c2}}{2}{(1+e^{i\delta_{2}{t}})}{\rho_{31}}-\Gamma_{3}{\rho_{33}}\nonumber\\
\dot{\rho}_{34}=&-\frac{i\Omega_\text{c2}^{\ast}}{2}{(1+e^{-i\delta_{2}{t}})}{\rho_{14}}-\gamma_{34}{\rho_{34}}\nonumber\\
\dot{\rho}_{44}=&\Gamma_{34}{\rho_{33}}+\Pi_{g}{(\rho_{11}-\rho_{44})}\nonumber
\end{align}
\normalsize{}

\section{Enhanced absorption for optical pumping system}
\label{AppendixB}

\subsection{Optical pumping system}\label{AppendB1}
\begin{figure}
   \begin{center}    
    \includegraphics[width =1.0\linewidth]{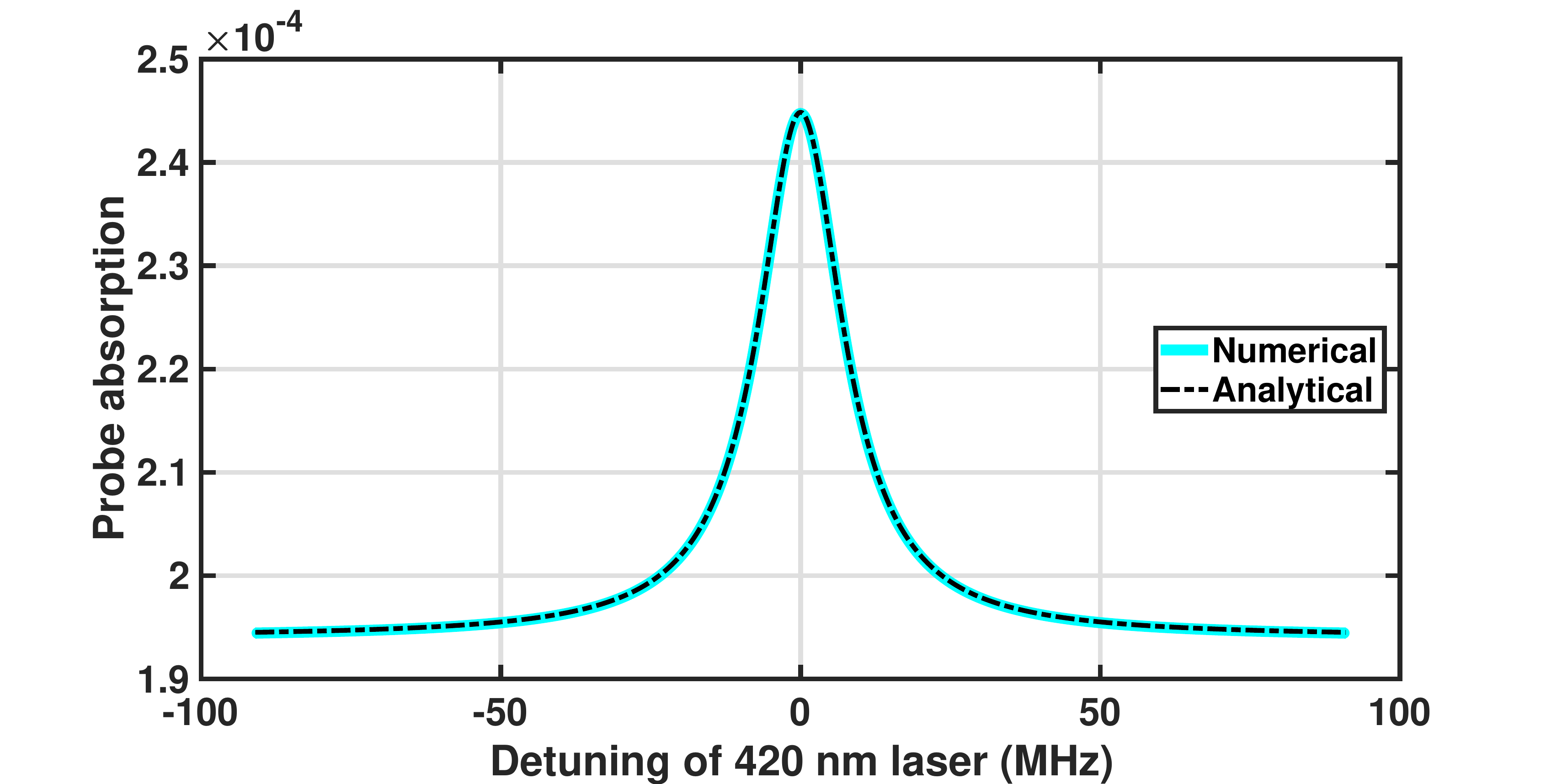}
      \caption{(Color online). Comparison of the full numerical solution of the density matrix in optical pumping system with the the analytical solution given in Eq. \ref{eqB3}. $\Omega_{c2}=\sqrt{1.5}\Gamma_{3}$, $\Gamma_{2}=2\pi\times6.065~\text{MHz},~\Gamma_{3}=2\pi\times1.32~\text{MHz}$.}
      \label{FigB1}
   \end{center}
\end{figure}

The Hamiltonian $\text{H}$ of the optical pumping system consider in Fig. \ref{Fig2}a under electric-dipole and rotating-wave approximation and in the interaction picture is given as follows,
\begin{align}
\label{eqB1}
\text{H}=&\frac{\hbar}{2}\big\{\Omega_{p}\ket{1}\bra{2}+\Omega_{c2}\ket{4}\bra{3}-\Delta_{p}\ket{2}\bra{2}-\Delta_{c2}\ket{3}\bra{3}+h.c.\big\}
\end{align}
The equations of motion of the density matrix is obtained from Eq. \ref{eq2} and \ref{eqB1} and set of equations are given as follows,
\begin{align}
\label{eqB2}
\dot{\rho}_{12}=&\frac{i\Omega_{p}}{2}{(\rho_{11}-\rho_{22})}-\gamma_{12}^{dec}\rho_{12}\\
\dot{\rho}_{13}=&-\frac{i\Omega_{p}}{2}{\rho_{23}}+\frac{i\Omega_{c2}^{\ast}}{2}{\rho_{14}}-\gamma_{13}\rho_{13}\nonumber\\
\dot{\rho}_{14}=&-\frac{i\Omega_{p}}{2}{\rho_{24}}+\frac{i\Omega_{c2}}{2}{\rho_{13}}-\gamma_{14}\rho_{14}\nonumber\\
\dot{\rho}_{22}=&-\frac{i\Omega_{p}^{\ast}}{2}{\rho_{12}}+\frac{i\Omega_{p}}{2}{\rho_{21}}-\Gamma_{2}{\rho_{22}}\nonumber\\
\dot{\rho}_{23}=&-\frac{i\Omega_{p}^{\ast}}{2}{\rho_{13}}-\gamma_{23}\rho_{23}+\frac{i\Omega_{c2}^{\ast}}{2}{\rho_{24}}\nonumber\\
\dot{\rho}_{24}=&-\frac{i\Omega_{p}^{\ast}}{2}{\rho_{14}}-\gamma_{24}\rho_{24}+\frac{i\Omega_{c2}}{2}{\rho_{23}}\nonumber\\
\dot{\rho}_{33}=&-\frac{i\Omega_{c2}}{2}{\rho_{43}}+\frac{i\Omega_{c2}^{\ast}}{2}{\rho_{34}}-\Gamma_{3}{\rho_{33}}\nonumber\\
\dot{\rho}_{34}=&-\frac{i\Omega_{c2}}{2}{(\rho_{33}-\rho_{44})}-\gamma_{34}{\rho_{34}}\nonumber\\
\dot{\rho}_{44}=&-\frac{i\Omega_{c2}^{\ast}}{2}{\rho_{34}}+\frac{i\Omega_{c2}}{2}{\rho_{43}}+\Gamma_{34}{\rho_{33}}+\Pi_{g}{(\rho_{11}-\rho_{44})}\nonumber
\end{align}
\normalsize{}where, $\gamma_{12}=i\Delta_{p}+\gamma^{dec}_{12}$, $\gamma_{13}=\gamma^{dec}_{13}$, $\gamma_{14}=i\Delta_{c2}+\gamma^{dec}_{14}$, $\gamma_{23}=-i\Delta_{p}+\gamma^{dec}_{23}$, $\gamma_{24}=i(\Delta_{c2}-\Delta_{p})+\gamma^{dec}_{24}$, $\gamma_{34}=i\Delta_{c2}+\gamma^{dec}_{34}$. The steady state solution of Eq. \ref{eqB2} in the weak probe approximation which gives enhanced absorption spectrum of the probe is expressed as follows,
\begin{widetext}
\begin{align}
\label{eqB3}
\rho_{12}=&\frac{\tabbedLongstack[l]{& i \Omega_{p} (\Omega_{c2}^2 (\Gamma_{3}+\Pi_{g}) (\Gamma_{31}+\Pi_{g})+\Gamma_{3} \Pi_{g} ((\Gamma_{3}+\Pi_{g})^2+4 \Delta_{c2}^2))}}{\tabbedLongstack[l]{&(\Gamma_{2}+\Pi_{g}+2 i \Delta_{p}) (\Omega_{c2}^2 (\Gamma_{3}+\Pi_{g}) (\Gamma_{31}+3 \Pi_{g})+2 \Gamma_{3} \Pi_{g} ((\Gamma_{3}+\Pi_{g})^2+4 \Delta_{c2}^2))}}
\end{align}
\end{widetext}
The solution of optical pumping system given in Eq. \ref{eqB3} is graphically represented in Fig. \ref{FigB1} and is well matched with the numerical simulation of the full density matrix given in Eq. \ref{eqB2}.
% Figure B1 %%%%%%%%%%%%%%%%%%%%%%%      
%%%%%%%%%%%%%%%%%%%%%%%%%%%%%%%%%% 
\subsection{VIPO at IR transition for optical pumping system}
\begin{figure}
   \begin{center}    
    \includegraphics[width =1.0\linewidth]{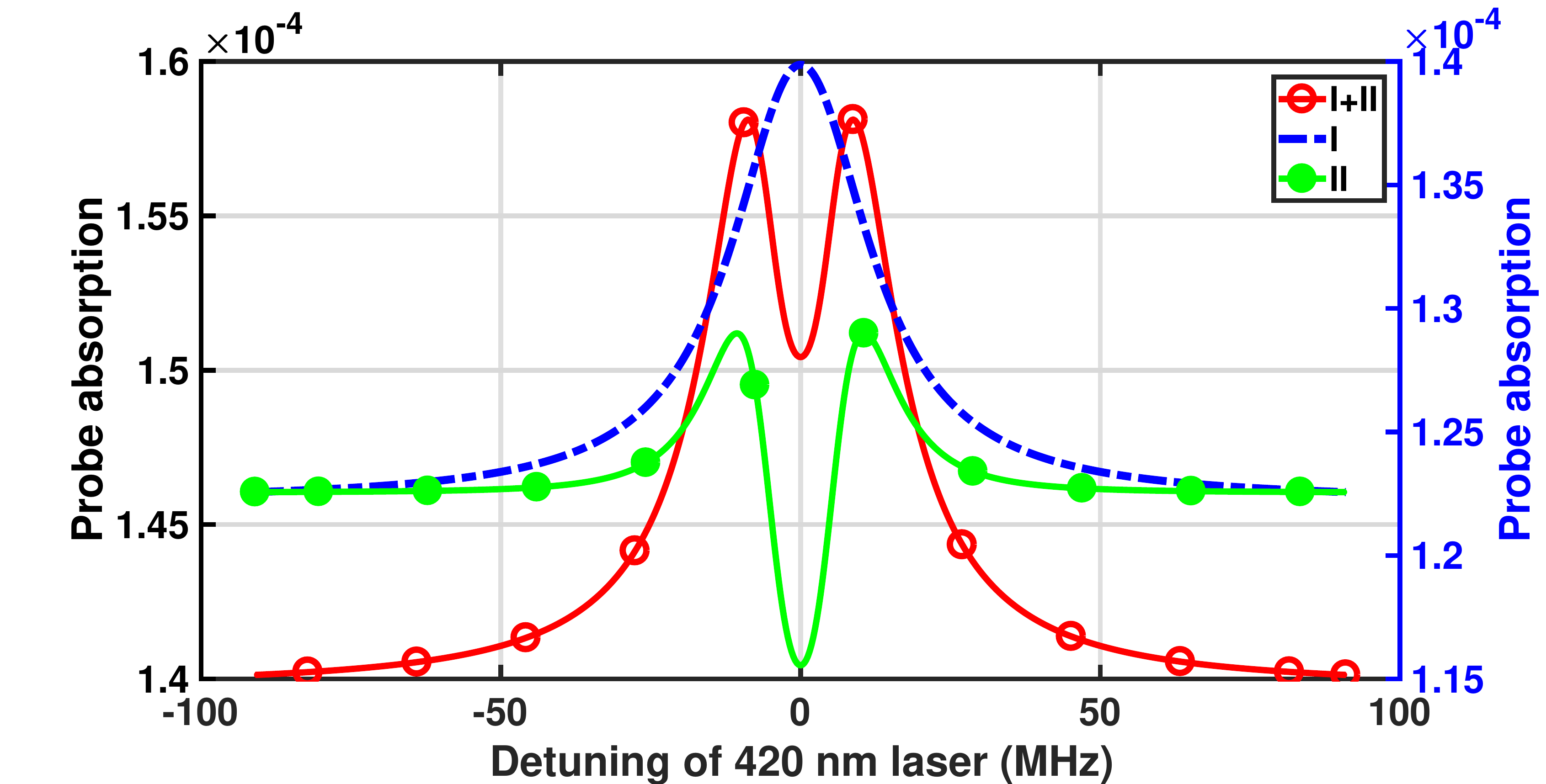}
      \caption{(Color online). The graphical representation of the individual terms I and II given in Eq. \ref{eq10}. $\Omega_{c1}=\Gamma_{2}$, $\Omega_{c2}=\sqrt{1.5}\Gamma_{3}$, $\Gamma_{2}=2\pi\times6.065~\text{MHz},~\Gamma_{3}=2\pi\times1.32~\text{MHz}$. The vertical axis of the red trace marked by circles is on the left and the blue dashed trace and the green trace marked with dots are on the right.}
      \label{FigB2}
   \end{center}
\end{figure}

The Hamiltonian of the VIPO at IR transition for the optical pumping system consider in Fig. \ref{Fig2}b is given in Eq. \ref{eq9}. The equations of motion of density matrix elements is also obtained from Eq. \ref{eq2} and \ref{eq9} which gives the following set of equations.\small{}
\begin{align}
\label{eqB4}
\dot{\rho}_{12}=&\frac{i}{2}{(\Omega_{c1}+\Omega_{p}e^{i\delta_{1}{t}})(\rho_{11}-\rho_{22})}-\gamma_{12}\rho_{12}\\
\dot{\rho}_{13}=&-\frac{i}{2}{(\Omega_{c1}+\Omega_{p}e^{i\delta_{1}{t}})}{\rho_{23}}+\frac{i\Omega_{c2}^{\ast}}{2}{\rho_{14}}-\gamma_{13}\rho_{13}\nonumber\\
\dot{\rho}_{14}=&-\frac{i}{2}{(\Omega_{c1}+\Omega_{p}e^{i\delta_{1}{t}})}{\rho_{24}}+\frac{i\Omega_{c2}}{2}{\rho_{13}}-\gamma_{14}\rho_{14}\nonumber\\
\dot{\rho}_{22}=&-\frac{i}{2}{(\Omega_{c1}^{\ast}+\Omega_{p}^{\ast}e^{-i\delta_{1}{t}})\rho_{12}}+\frac{i}{2}{(\Omega_{c1}+\Omega_{p}e^{i\delta_{1}{t}})\rho_{21}}-\Gamma_{2}{\rho_{22}}\nonumber\\
\dot{\rho}_{23}=&-\frac{i}{2}{(\Omega_{c1}^{\ast}+\Omega_{p}^{\ast}e^{-i\delta_{1}{t}})\rho_{13}}-\gamma_{23}\rho_{23}+\frac{i\Omega_{c2}^{\ast}}{2}{\rho_{24}}\nonumber\\
\dot{\rho}_{24}=&-\frac{i}{2}{(\Omega_{c1}^{\ast}+\Omega_{p}^{\ast}e^{-i\delta_{1}{t}})\rho_{14}}-\gamma_{24}\rho_{24}+\frac{i\Omega_{c2}}{2}{\rho_{23}}\nonumber\\
\dot{\rho}_{33}=&-\frac{i\Omega_{c2}}{2}{\rho_{43}}+\frac{i\Omega_{c2}^{\ast}}{2}{\rho_{34}}-\Gamma_{3}{\rho_{33}}\nonumber\\
\dot{\rho}_{34}=&-\frac{i\Omega_{c2}}{2}{(\rho_{33}-\rho_{44})}-\gamma_{34}{\rho_{34}}\nonumber\\
\dot{\rho}_{44}=&-\frac{i\Omega_{c2}^{\ast}}{2}{\rho_{34}}+\frac{i\Omega_{c2}}{2}{\rho_{43}}+\Gamma_{34}{\rho_{33}}+\Pi_{g}{(\rho_{11}-\rho_{44})}\nonumber
\end{align}
\normalsize{}
The steady state solution of the equations of motion given in Eq. \ref{eqB4} is given in Eq. \ref{eq10} and the solution of the various density matrix components $\rho_{11}^{(0)}$, $\rho_{22}^{(0)}$, $\rho_{11}^{(+1)}$ and $\rho_{22}^{(+1)}$ is given as follows. The individual contribution of the terms I and II given in Eq. \ref{eq10} is also illustrated in Fig. \ref{FigB2}.
% Figure B2 %%%%%%%%%%%%%%%%%%%%%%%      
%%%%%%%%%%%%%%%%%%%%%%%%%%%%%%%%%% 

\subsection{VIPO at IR and VSS at blue transition for optical pumping system}

The Hamiltonian of the VIPO at IR transition and VSS at blue transition for the optical pumping system consider in Fig. \ref{Fig2}c is given in Eq. \ref{eq11}. The equations of motion of the density matrix is obtained from Eq. \ref{eq2} and \ref{eq11} as follows.\small{}
\begin{align}
\label{eqB5}
\dot{\rho}_{12}=&\frac{i}{2}{(\Omega_{c1}+\Omega_{p}e^{i\delta_{1}{t}})(\rho_{11}-\rho_{22})}-\gamma_{12}^{dec}\rho_{12}\\
\dot{\rho}_{13}=&-\frac{i}{2}{(\Omega_{c1}+\Omega_{p}e^{i\delta_{1}{t}})}{\rho_{23}}+\frac{i\Omega_{c2}^{\ast}}{2}{(1+e^{-i\delta_{2}{t}})}{\rho_{14}}-\gamma_{13}\rho_{13}\nonumber\\
\dot{\rho}_{14}=&-\frac{i}{2}{(\Omega_{c1}+\Omega_{p}e^{i\delta_{1}{t}})}{\rho_{24}}+\frac{i\Omega_{c2}}{2}{(1+e^{i\delta_{2}{t}})}{\rho_{13}}-\gamma_{14}\rho_{14}\nonumber\\
\dot{\rho}_{22}=&-\frac{i}{2}{(\Omega_{c1}^{\ast}+\Omega_{p}^{\ast}e^{-i\delta_{1}{t}})\rho_{12}}+\frac{i}{2}{(\Omega_{c1}+\Omega_{p}e^{i\delta_{1}{t}})\rho_{21}}-\Gamma_{2}{\rho_{22}}\nonumber\\
\dot{\rho}_{23}=&-\frac{i}{2}{(\Omega_{c1}^{\ast}+\Omega_{p}^{\ast}e^{-i\delta_{1}{t}})\rho_{13}}+\frac{i\Omega_{c2}^{\ast}}{2}{(1+e^{-i\delta_{2}{t}})}{\rho_{24}}-\gamma_{23}\rho_{23}\nonumber\\
\dot{\rho}_{24}=&-\frac{i}{2}{(\Omega_{c1}^{\ast}+\Omega_{p}^{\ast}e^{-i\delta_{1}{t}})\rho_{14}}+\frac{i\Omega_{c2}}{2}{(1+e^{i\delta_{2}{t}})}{\rho_{23}}-\gamma_{24}\rho_{24}\nonumber\\
\dot{\rho}_{33}=&-\frac{i\Omega_{c2}}{2}{(1+e^{i\delta_{2}{t}})}{\rho_{43}}+\frac{i\Omega_{c2}^{\ast}}{2}{(1+e^{-i\delta_{2}{t}})}{\rho_{34}}-\Gamma_{3}{\rho_{33}}\nonumber\\
\dot{\rho}_{34}=&-\frac{i\Omega_{c2}}{2}{(1+e^{i\delta_{2}{t}})}{(\rho_{33}-\rho_{44})}-\gamma_{34}{\rho_{34}}\nonumber\\
\dot{\rho}_{44}=&-\frac{i\Omega_{c2}^{\ast}}{2}{(1+e^{-i\delta_{2}{t}})}{\rho_{34}}+\frac{i\Omega_{c2}}{2}{(1+e^{i\delta_{2}{t}})}{\rho_{43}}+\Gamma_{34}{\rho_{33}}\nonumber\\
&+\Pi_{g}{(\rho_{11}-\rho_{44})}\nonumber
\end{align}

\bibliography{CPOSpaper}

\end{document}